\documentclass[12pt]{iopart}

\pdfoutput=1

\usepackage{graphicx}
\usepackage{iopams}
\usepackage{url}

\newcommand{\bra}[1]{\langle{#1}|}
\newcommand{\braket}[2]{\langle{#1}|{#2}\rangle}
\newcommand{\ket}[1]{|{#1}\rangle}
\newcommand{\upd}{\mathrm{d}}
\newcommand{\df}[1]{\delta\!\left({#1}\right)}

\begin{document}

\title[On the concentration of large deviations for fat tailed distributions]{On the concentration of large deviations for fat tailed distributions, with application to financial data}

\author{Mario Filiasi$^{1,2}$, Giacomo Livan$^3$, Matteo Marsili$^3$, Maria Peressi$^1$, Erik Vesselli$^{1}$, Elia Zarinelli$^{2}$}
\address{${}^{1}$Physics Department, University of Trieste, via Valerio 2, I-34127 - Trieste, Italy\\
${}^{2}$LIST S.p.A., via Carducci 20, I-34122 - Trieste, Italy\\
${}^{3}$The Abdus Salam International Centre for Theoretical Physics (ICTP), Strada Costiera 11, I-34014 - Trieste, Italy\\
}

\begin{abstract}
Large deviations for fat tailed distributions, i.e. those that decay slower than exponential, are not only relatively likely, but they also occur in a rather peculiar way where a finite fraction of the whole sample deviation is concentrated on a single variable. The regime of large deviations is separated from the regime of typical fluctuations by a phase transition where the symmetry between the points in the sample is {\em spontaneously broken}.
For stochastic processes with a fat tailed microscopic noise, this implies that while typical realizations are well described by a diffusion process with continuous sample paths, large deviation paths are typically discontinuous. 
For eigenvalues of random matrices with fat tailed distributed elements, a large deviation where the trace of the matrix is anomalously large concentrates on just a single eigenvalue, whereas in the thin tailed world the large deviation affects the whole distribution. 

These results find a natural application to finance. Since the price dynamics of financial stocks is characterized by fat tailed increments, large fluctuations of stock prices are expected to be realized by discrete jumps. Interestingly,
we find that large excursions of prices are more likely realized by continuous drifts rather than by discontinuous jumps.
Indeed, auto-correlations suppress the concentration of large deviations. Financial covariance matrices also exhibit an anomalously large eigenvalue, the market mode, as compared to the prediction of random matrix theory. We show that this is explained by a large deviation with excess covariance rather than by one with excess volatility. \end{abstract}

\maketitle

Large deviations are rare events where sample averages do not take their typical values, i.e. the expected value, but rather deviate systematically from it. 
Large Deviation Theory (LDT) \cite{ellis,varadhan} has been developed to assess the probability of rare sample fluctuations, and it is  a central subject in statistical physics \cite{Touchette}, information theory, statistical inference and learning \cite{Cov,MM}. LDT shows that when the distribution of the variables decays at least exponentially fast in the tails, large deviations are exponentially rare in the size of the sample,
and they are realized as independent draws from a modified distribution.

The extension of LDT to distributions with fat tails, i.e. those that decay slower than an exponential, have been discussed in the context of mass transport models in physics \cite{Burda,ZRP,satya,satya_2} and of extremal events in risk theory \cite{Embrechts}. In this case, large deviations not only become much more likely, but the way in which they are realized is rather peculiar. 
The whole excess of the sample deviation with respect to the typical value concentrates on a single variable. Loosely speaking, while large deviations result from the accumulation of many small deviations in the standard case of thin tailed distributions, for fat tailed ones large deviations arise as a consequence of a huge fluctuation in a single term.

The regime where large deviations concentrate has all the properties of a phase with a {\em spontaneously broken} symmetry, which is the permutation symmetry between the points in the sample. 
For distributions with fat tails only on one side, this phase is separated from the regime where large deviations are realized in a symmetric fashion by a singularity in the second derivative of the Cramer rate function, i.e. of the entropy. 
This is perhaps the simplest realization of a second order phase transition with spontaneous symmetry breaking. 

In what follows, we first review LDT for independent variables and its extension to fat tailed distributions, drawing on the literature on mass transport models \cite{Burda,ZRP,satya,satya_2} and risk modeling  \cite{Embrechts}. We then discuss the extension of these findings to non independent variables. The example of elliptic distributions shows that positive correlations are expected to suppress the concentration of large deviations. On the contrary, concentration persists for eigenvalues of random matrices, that are known to ``repel'' each other. There, we show that both in the Wigner \cite{Wigner} and the Wishart \cite{Wishart} ensemble, a large deviation in the average eigenvalue -- or in the trace of the matrix -- concentrates on the largest eigenvalue when the underlying distribution is fat tailed. This contrasts with what happens for thin tailed distribution where the large deviation affects the whole distribution -- though in different ways.

The second part of the paper revisits the statistics of returns of financial stocks on the basis of these insights. First we analyze large fluctuations in the returns of financial stocks. This relates to large deviation paths for stochastic processes. It is well known that if the increments -- the log-returns in our case -- have a finite variance, typical realizations are well described by a diffusion process with continuous sample paths. Yet, if increments have a fat tailed distribution, 
paths with an anomalously large excursion are expected to be discontinuous. 
This markedly differs from the behavior of LDT for diffusion processes, that are realized by continuous sample paths. 

In financial time series, whose increment are fat tailed distributed with a finite variance \cite{Stanley}, large discontinuous movements in prices -- like the so called {\em flash crashes} -- have attracted considerable attention \cite{Kirilenko,Bormetti}.
Yet we find that large returns are much less concentrated than expected.

Random matrix theory has also been widely used in order to describe the covariance of financial time series \cite{Laloux,Plerou,Plerou2}. Financial covariance matrices differ markedly from sample covariance matrices of random walks, that are well described by the Wishart ensemble \cite{Wishart}, because of the presence of a very large eigenvalue, the so-called market mode. The origin of the market mode can be traced back \cite{RaffaelliMarsiliPonsot} to widespread portfolio optimization strategies and the Capital Asset Pricing Model (CAPM) \cite{CAPM}. Within the present discussion, it is temping to relate it to a large deviation in random matrix ensemble with fat tailed microscopic noise.
Large deviations of the trace of the covariance matrix translate in the statement that the volatility of stocks is larger than expected. As we show, such excess volatility concentrate in a very large eigenvalue, whose eigenvector is localized both on one stock and on one time point. By contrast, the market mode has an extended eigenvector both across the stocks and in time. We show that the emergence of a market mode is consistent with a large deviation where the total covariance is larger than expected. Excess covariance indeed reproduces both a maximal eigenvalue that carries a large fraction of the whole trace and an extended market mode, both across stocks and in time. The last section concludes with a discussion on possible further applications of these results.

\section{Large deviations for i.i.d. variables with thin tailed distribution}
\label{sec:thintails}

Consider a sample $\underline{X}=\{X_1,\ldots, X_N\}$, where $X_i\in \mathcal{X}$ are i.i.d.\
\footnote{Notations: i.i.d.\ stands for independent and identically distributed. We use capital letters $Q(x)=P\{X\le x\}$ for cumulative distribution functions (c.d.f.) and the corresponding lowercase $q(x)=Q'(x)$ for probability density functions (p.d.f.). When the random variable $X$ has finite support $\mathcal{X}$, then $q(x)=P\{X=x\}$ denotes the probability mass function. The expected value over a distribution $Q$ is denoted as $\langle\ldots\rangle_Q$.}\
draws from a distribution $Q(x)=P\{X_i\le x\}$. 
We assume that both $\langle X\rangle_Q$ and $\langle X^2\rangle_Q$ are finite and that $N$ is large.
We are interested in estimating the probability of events
\begin{equation*}
E=\left\{ \frac{1}{N}\sum_{i=1}^N X_i \in [\bar x,\bar x+\delta x]\right\} \ .
\end{equation*}
The Law of Large Numbers  \cite{Gned}, implies that for any $\epsilon>0$
\begin{equation*}	
\lim_{N\to\infty} P\left\{\left|\frac{1}{N}\sum_{i=1}^N X_i-\langle X\rangle_Q\right|>\epsilon\right\}=0 \ .
\end{equation*}
So, if $\langle X\rangle_Q\not\in [\bar x,\bar x+\delta x]$ the event $E$ is not typical, i.e. it has a vanishing probability as $N\to\infty$. 

In the case where $X_i$ take value in a finite set $\mathcal{X}$ 
we can invoke Sanov's theorem \cite{Cov,MM} as follows. Let us denote by $P(x)$ a generic distribution on $\mathcal{X}$ such that $E$ is typical. The difference between $P(x)$ and the true distribution $Q(x)$ is naturally measured in term of the Kullback-Leibler divergence $D_{KL}(P||Q)=\sum_{x\in\mathcal{X}} p(x)\log[p(x)/q(x)]$, where $p(x)$ and $q(x)$ are the probability mass functions associated to $P(x)$ and $Q(x)$, respectively. Sanov's theorem states that the probability of $E$ is asymptotically given by
\[ P\{E\}\simeq e^{-N D_{KL}(P^*||Q)} \ , \]
where $P^*(x)$ is the distribution that minimizes $D_{KL}(P||Q)$ over all possible $P$'s. LDT then boils down to a problem of constrained optimization, that is solved by introducing the constraint $\sum_i X_i/N\in [\bar x,\bar x+\delta x]$ with a Lagrange multiplier in the optimization of $D_{KL}(P||Q)$. 
The result is given by
\begin{equation}
\label{Pstar}
p^*(x)=p_\beta(x)\equiv \frac{1}{Z(\beta)} q(x) e^{-\beta x} \ ,
\end{equation}
where $Z(\beta)$ is a normalization constant and $\beta$ is adjusted so that $\langle X\rangle_{P_\beta}=\sum_{x\in\mathcal{X}}
p_\beta (x) x=\bar x$. These considerations generalize to the case when $\mathcal{X}$ is a continuous set and $q(x)$ has either finite support or decays at least as fast as an exponential for $|x|\to\infty$. In both cases, the probability of $E$ for infinitesimal $\delta x$, is expressed in terms of the p.d.f.\ of $\bar x$ as $P\{E\}=\rho_N(\bar x)\delta x\propto e^{-NI(\bar x)}\delta x$, where the rate function $I(\bar x)$ (also called Cramer's function) is given by
\begin{equation}
I(\bar x)=-\lim_{N\to\infty}\frac{1}{N}\log P\{E\} =D_{KL}(P_\beta||Q) \ ,
\label{Cramer}
\end{equation}
where $P_\beta$ is obtained from Eq.\ (\ref{Pstar}) and $\beta$ is such that $\langle X\rangle_{P_\beta}=\bar x$.
In practice, within these assumptions, the rate function $I(\bar x)$ for i.i.d.\ samples can be computed from the function $\phi(h)=\log\langle e^{hX}\rangle_Q$ through the Legendre transform \cite{Cov,MM} (with $h=-\beta$).

It's worth to recall the derivation of this result within the density functional formalism, as explained in Ref. \cite{MM}. 
This will be useful to address more complex cases below.
For infinitesimal $\delta x$, we have $P\{E\}=\rho_N(\bar{x})\delta x$ with
\begin{equation}
\rho_N(\bar{x}) = \int \left(\prod_{i=1}^N \upd x_i\,q(x_i) \right) \df{\frac{1}{N}\sum_{i=1}^N x_i - \bar{x}} \ .
\label{PDF_START}
\end{equation}
In the limit $N\to\infty$ we can exchange the multiple integral over the $x$'s with a unique functional integral over the density $p(x) = \frac{1}{N}\sum_{i=1}^N\df{X_i-x} $. We stress that this corresponds to making a {\em symmetric ansatz} where we assume that all variables are drawn from the same distribution. This leads to 
\[ \rho_N(\bar{x}) \approx \int \mathcal{D}p(x)\,e^{-ND_{KL}(P||Q)}\,\df{\int\upd x\, p(x) - 1}\df{\int\upd x\,x\,p(x) - \bar{x}} \ , \]
where the additional delta function fixes the correct normalization of $p(x)$. For large $N$, the integral can be evaluated by means of the saddle point approximation. This yields again Eq.\ (\ref{Pstar}), where the saddle point solution $p^*(x)$ minimizes the Kullback-Leibler divergence under the constraints expressed by the delta functions. The solution, thus, matches exactly with Eq.\ (\ref{Pstar}) and leads to:
\[ I(\bar{x}) = D_{KL}(P^*||Q) = h\bar{x} - \phi(h) \ , \]
where $h=-\beta$ and $\phi(h)=\log Z(\beta)$. Finally, re-expressing the constraint $\langle X\rangle_{P_\beta}=\bar x$ in the form $\frac{\upd}{\upd h}\phi(h)=\bar{x}$, one finds that $I(\bar{x})$ is the Legendre transform of $\phi(h)$.

This procedure not only determines the probability of the large deviation $E$, but 
it also informs us of how untypical outcomes are ``typically'' realized: one can show \cite{Cov} that the distribution of $X$, conditional on the occurrence of $E$, is given by $P\{X\le x|E\}=P_\beta(x)$. Thus a sample $\underline X$ exhibiting a large deviation in the mean $\bar x=\sum_iX_i/N  \neq \langle X\rangle_Q$ can be thought of as a sequence drawn independently from $P_\beta(x)$ in Eq. (\ref{Pstar}).

The rate function $I(\bar x)$ has the property that it is positive and it vanishes for $\bar x=\langle X\rangle_Q$, which corresponds to the point $h=\beta=0$. This description holds if $\phi(h)$ exists at least for $h$ in an open neighborhood of the origin, i.e. if the p.d.f.\ of $X$ decays at least as an exponential for $|x|\to\infty$. What happens if this is not true? 

\section{Fat tailed distributions}
\label{sec:fattails}

A fat tailed distribution is any distribution $Q(x)$ such that $e^{hx}q(x)$ diverges for all $h>0$, when $x\to\infty$. For simplicity, we focus on the right tail of the p.d.f., and assume that $q(x)$ vanishes at least exponentially fast as $x\to-\infty$. 
This includes stretched exponential distributions $q(x)\sim e^{-a x^\alpha}$ with $\alpha<1$ and power law distributions $q(x)\sim A x^{-\gamma}$ with $\gamma>1$. In the following we assume that the distribution $Q(x)$ has finite mean and variance, so that we can invoke both the Law of Large Numbers and the Central Limit Theorem (in case of a power law distribution, this implies $\gamma>3$).

\subsection{Large deviations and phase transitions}

It is instructive to consider first an example, borrowed from Ref. \cite{Burda}, that provides the basic intuition on the behavior of large deviations for fat tailed distributions and on its relation with second order phase transitions. Consider the specific case $q(x)=A/(1+x)^\gamma$ for $x\in \mathbb{N}$, where $A=1/\zeta(\gamma)$ and $\gamma>3$.
It is clear that the normalizing constant
\[ Z(\beta)=\sum_{k=0}^\infty \frac{Ae^{-\beta k}}{(1+k)^ \gamma} \]
in Eq. (\ref{Pstar}) is finite only for $\beta\ge 0$. The expected value of $X$ under the distribution $P_\beta(x)$
\[
\langle X\rangle_{P_\beta}=-\frac{d}{d\beta}\log Z(\beta)
\]
is a decreasing function of $\beta$ and $\langle X\rangle_{P_{\beta=0}}=\langle X\rangle_Q$. Hence, the recipe for large deviations works for all deviations where $\bar x\le \langle X\rangle_Q$ as it is always possible to find a value of $\beta(\bar x)$ such that $\langle X\rangle_{P_\beta}=\bar x$. In words, it is always possible to introduce a (exponential) cutoff to the distribution of $X$ in order to reduce its expected value (see also \cite{didierbook} sect. 3.3.5). 

What about large deviations with $\bar x>\langle X\rangle_Q$? It is easy to show that there are typical ways to realize large fluctuations where the excess of the average is taken up by a single variable. Indeed, consider all the samples $\underline X$ such that the sum over all but the largest variable $X_{i^*}$ is typical, i.e. 
\begin{equation}
\label{eventistar}
\left|\frac{1}{N-1}\sum_{i\neq i^*}X_i- \langle X\rangle_Q\right|<\epsilon \ ,
\end{equation}
and $X_{i^*}\cong N(\bar x -\langle X\rangle_Q)+\langle X\rangle_Q$. For each such sample the average takes the value $\bar x$, hence the probability to observe $E$ is at least
\[
P\{E\}\ge N q(X_{i^*})=AN\left[N(\bar x -\langle X\rangle)+\langle X\rangle\right]^{-\gamma}
\]
because the event in Eq. (\ref{eventistar}) occurs with probability arbitrarily close to one, and there are $N$ ways in which $i^*$ can be chosen. This means that, if we 
define the rate function $I(\bar x)$ as in Eq. (\ref{Cramer}), then $I(\bar x)=0$ for all $\bar x\ge \langle X\rangle_Q$. We recall that, for 
$\bar x< \langle X\rangle_Q$, the Cramer function is well approximated by a quadratic function $I(\bar x)\simeq \frac{a}{2}\left(\bar x-\langle X\rangle_Q\right)$ in the left neighborhood of $\langle X\rangle_Q$. Therefore $I(\bar x)$ has a discontinuity in its second derivative that jumps from $a>0$ to zero at $\bar x = \langle X\rangle_Q$. This singularity is accompanied by the {\em spontanous breaking of the symmetry} between the variables $X_i$, because one of them takes an {\em extensive} value (i.e. a value proportional to $N$). This is precisely the phenomenology of second order phase transition in statistical physics. Notice that the critical point corresponds to the point where the average  takes its typical value $\bar x=\langle X\rangle_Q$, which suggests that in a precise sense, systems characterized by fat tailed distributions are poised at a critical point.

The same phenomenon of the concentration of large deviations occurs for stretched exponential distribution or for log-normal distributions, as we now show within the density functional formalism.

\subsection{Derivation in the general case}
\label{sec:funcint_cond}

In order to characterize large deviations in fat tailed distribution we can exploit again the density functional formalism (see \cite{MM} and previous section), by means of a simple \emph{concentration ansatz}. The ansatz relies on the assumption that the symmetry of the system -- the invariance under exchange of variables -- is \emph{spontaneously broken}, and so that the variables can obey different scaling laws. Our method is the following. We write $P\{E\}=\rho_N(\bar{x})\delta x$, where $\rho_N(\bar{x})$ is given by Eq.\ (\ref{PDF_START}), then we express again $\rho_N(\bar{x})$ as a functional integral over some density $p(x)$. Unlike the previous case, {\em symmetry breaking} enters in the fact that the density $p(x)$ describes all variables but one, namely $X_{i^*}$. At the leading order in $N$, this leads to:
\begin{eqnarray*}
\rho_N(\bar{x}) &\approx& \int \mathcal{D}p(x)\,\upd x_{i^*}\,e^{-F(P,Q)}\,\df{\int\upd x\, p(x) - 1} \times \\ \nonumber
&\times& \df{\frac{N-1}{N}\int\upd x\,x\,p(x) +\frac{x_{i^*}}{N}- \bar{x}} \ ,
\end{eqnarray*}
where:
\begin{equation}
F(P,Q) = (N-1) D_{KL}(P||Q) - \ln q(x_{i^*}) \ .
\label{EFF_ENERGY}
\end{equation}
In order to evaluate this integral through the saddle-point approximation, we must rescale $p(x)$, $x_{i^*}$, and $\bar{x}$ such that all leading terms in the integral are of the same order in $N$. The correct scaling laws are fixed by the constraints in the delta functions, and the only non-trivial scaling is given by $x_{i^*}\sim N$. For this reason, we perform the substitution $x_{i^*} = Nt$ and neglect all sub-leading terms for $N\to\infty$. Since $q(x)$ is fat tailed, the last term in Eq.\ (\ref{EFF_ENERGY}) is sub-linear and can be neglected, so we get:
\begin{eqnarray*}
\rho_N(\bar{x}) &\approx& \int \mathcal{D}p(x)\,\upd t\,e^{-ND_{KL}(P||Q)}\,\df{\int\upd x\, p(x) - 1} \times \\ \nonumber 
&\times&\df{\int\upd x\, x p(x) +t- \bar{x}} \ .
\end{eqnarray*}
We are now ready to perform the saddle point approximation. The minimization of the Kullback-Leibler divergence under the constraints in the delta function can be easily performed through the method of Lagrange multipliers. The new saddle-point solution reads:
\begin{equation}
p^*(x)=q(x) \qquad \mathrm{and}\qquad t^* = \bar{x}-\langle X\rangle_Q
\label{ORIGINAL_DISTR}
\end{equation}
This solution causes $D_{KL}(P||Q)$ to vanish and does not allow us to estimate the probability $P\{E\}$ (this is not surprising since, as we already argued, $I(\bar{x})=0$ and $P\{E\}$ is not exponentially suppressed in $N$). Yet, the above solution describes how large deviations in fat tailed distribution are typically realized. Eq.\ (\ref{ORIGINAL_DISTR}) shows that, in the limit $N\to\infty$, the variables tend to be distributed according to their original distribution $q(x)$, so they behave typically and are not sensitive to the fact that a large deviation is occurring. This is true for all variables but one, $X_{i^*}$, which scales as $X_{i^*}\simeq Nt^*=N(\bar{x}-\langle X\rangle_Q)$ and is the unique responsible for the whole deviation.

Summarizing, for fat tailed distributions {\em i)} large deviations are typically realized by {\em breaking the symmetry} between the variables $X_i$ and having one of them take an {\em extensive} value (i.e. a value proportional to $N$). {\em ii)} 
The fact that $I(\bar x)=0$ for all $\bar x\ge \langle X\rangle$ implies that $I(x)$ has a singularity at $x=\langle X\rangle_Q$ in the second derivative. As pointed out above, this is exactly what happens in a second order phase transition in statistical physics. 

\subsection{Implications for inference}

The different behavior of large deviations for fat and thin tails has important consequences on what we learn from observations. Imagine you have a theory based on the following assumption on a system of $N$ variables $X_i$:
\begin{description}
  \item[$\mathbf{A_1}$] $X_1,\ldots,X_N$ are i.i.d. 
  \item[$\mathbf{A_2}$] $X_i$ are drawn from a distribution $P\{X_i\le x\}=Q(x)$ with finite second moment
\end{description}
On the basis of this, the theory predicts that $\bar x=\frac{1}{N}\sum_{i=1}^N X_i\simeq \langle X\rangle_Q$. 
Imagine now that you observe an average $\bar x$ markedly different from $\langle X\rangle_Q$. What should you conclude?

In the case of thin tails, you would conclude that $A_2$ is false, i.e. that the variables $X_i$ are drawn from a modified distribution $P_\beta\not = Q$. In the case of fat tails, instead, you should instead question $A_1$: most likely all but one variable are drawn from the same distribution $Q(x)$ and there is a ``special'' variable that alone explains the observation of $\bar x$.

\subsection{Relation to condensation phase transition in mass transport models}

The resemblance of the concentration of large deviations for fat tailed distribution with a {\em phase transition} in statistical physics is made 
explicit by the mapping of the large deviation setup discussed above into the statistical properties of a gas of interacting particles, as discussed e.g. in \cite{Burda,satya,satya_2}. In brief, consider $\rho N$ particles distributed on $N$ sites with a Hamiltonian 
$H\{\underline n\}=\sum_{i=1}^N h(n_i)$ that allows only the $n_i$ particles on the same site $i$ to interact. At temperature $T$ one expects a factorized stationary state $Q\{n_i\}\propto \prod_i e^{-h(n_i)/T}$. Yet the total number of particles is fixed, which implies that $\sum_i n_i=N\rho$. This is usually achieved by introducing a chemical potential $\mu$, i.e. modifying the distribution to $P_\mu\{n_i\}\propto \prod_i e^{-h(n_i)/T-\mu n_i/T}$ and fixing $\mu$ so that $\langle n_i\rangle_{P_\mu}=\rho$. Yet, if $\rho>\langle n_i\rangle_Q$ this requires a negative value of $\mu$ and if $h(n)$ grows slower than linearly when $n\to\infty$, then this strategy is not applicable because $P_\mu$ is not normalizable. In such a case, the excess fraction $\rho - \langle n_i\rangle_Q$ of the particles {\em condensate} on a single site. This phenomenon  occurs not only in equilibrium systems but also in non-equilibrium systems of interacting particles that admit a factorizable stationary state \cite{satya,satya_2}. Indeed it is precisely the same mechanism of concentration of large deviations that we have discussed above. 

Majumdar, Evans and Zia \cite{satya,satya_2} derive exact and asymptotic results for the rate function and for the p.d.f.\ of the mass $n_i$. Their results provide a detailed description of LDT for i.i.d.\ variables with fat tailed distribution, to which we refer the interested reader.\
\footnote{Besides describing the case of different distributions, Refs.\ \cite{satya,satya_2} also discuss the case where the Law of Large Numbers and the Central Limit Theorem do not hold, i.e.\ $Q(x)\sim Ax^{-\gamma}$ with $1<\gamma\leq 3$.
It can be shown that the condensation phase transition occurs also for $2<\gamma\leq 3$, but the condensed phase is characterized by anomalous (i.e\ non-Gaussian) fluctuations. On the contrary, no phase transition takes place in the range $1<\gamma\leq 2$. In the latter case, indeed, the sum of random variables can be well approximated by the largest terms \cite{Zaliapin}, and (pseudo) condensation typically occurs because large deviations are spontaneous, as discussed also in \cite{BouchaudMezard}.}
The distribution of the maximum in the sample also exhibit interesting properties, for which we refer to \cite{evans}. Frisch and Sornette \cite{sf} report a similar symmetry breaking phenomenon for stretched exponential distributed variables, in the regime of {\em extreme} deviations, i.e.\ when $\bar x\to\infty$ with $N$ large but finite. 

\subsection{Large excursions in stochastic processes}
\label{sec:stocproc}

The concentration of large deviations has a striking manifestation in stochastic processes with independent increments. Consider the sum
\[
S_n=X_1+X_2+\ldots+X_n \ ,
\]
where $X_i$ are i.i.d. random variables with zero mean and finite variance $\sigma^2=\langle X^2_i\rangle_Q$. Taking $N\gg 1$, it is possible to define a continuous time process $W_t=S_n/\sqrt{N}$ in the rescaled time $t=n/N$. The statistics of $W_t$ is well described by the Wiener process, that has continuous sample paths. The conditional probability to find $W_t=w$ given that $W_{t=0}=0$ is given by the
Gaussian kernel
\[
p(w,t|0,0)=\frac{e^{-w^2/(2t)}}{\sqrt{2\pi t}}.
\]
Large deviations where $W_t-W_0=v(t-t_0)$ that is much larger than the typical excursion  $\sim \sqrt{t-t_0}$ are again described by the Gaussian kernel of a Wiener process with a drift
\[
p(w,t|0,0)=\frac{e^{-(w-vt)^2/(2t)}}{\sqrt{2\pi t}}.
\]
This again has continuous sample paths. 

\begin{figure}
\centering
\includegraphics[width=4in]{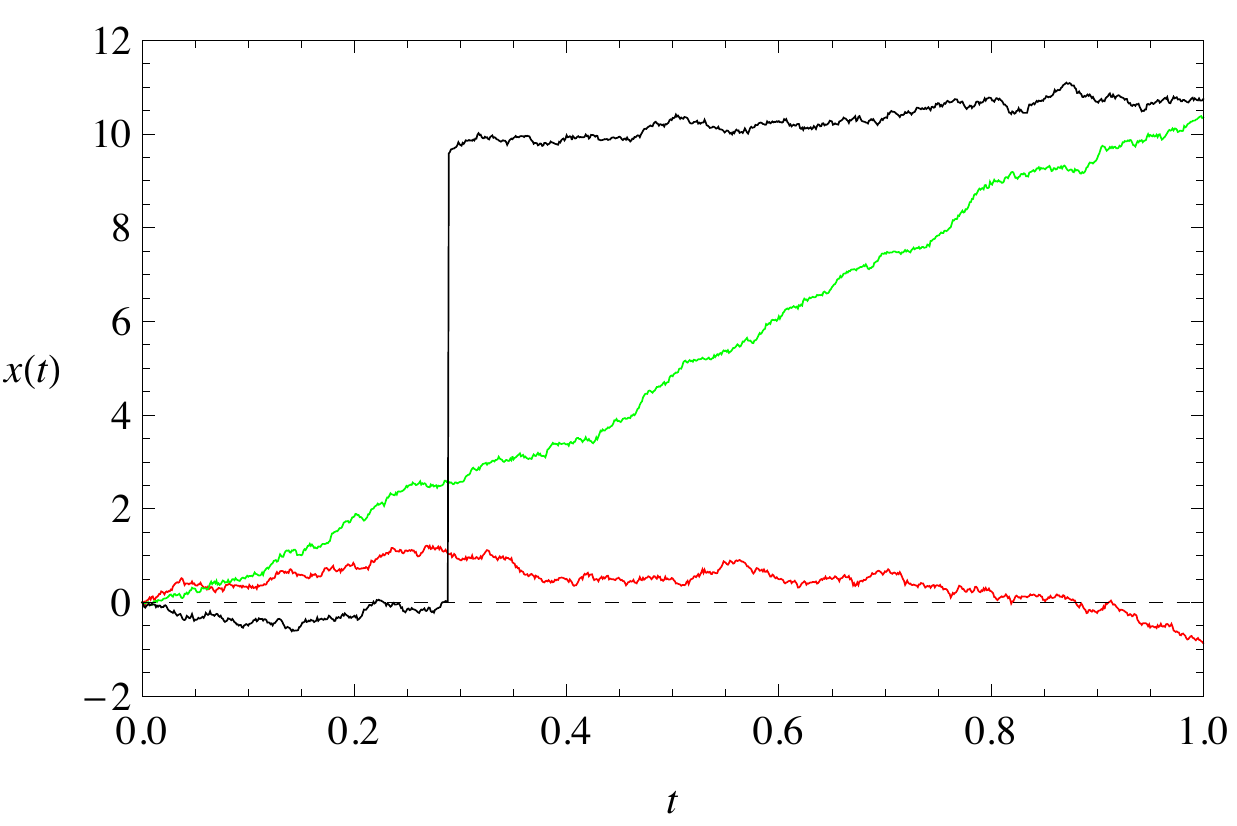}
\caption{Schematic illustration of a typical trajectory of a random walk (red line) and of large deviation paths with thin tailed distributed increments (green) and with fat tailed ones (black).}
\label{FIG:RWLD}
\end{figure}

However, if the microscopic increments have a fat tailed distribution, i.e.\ if their p.d.f.\ $p(x)$ decays to infinity slower than exponentially, while typical paths are continuous sample paths, 
we should expect that large excursions are realized as discontinuous paths with jumps. More precisely, one of the increments $X_i$ will account for the whole excursion. This phenomenon is schematically illustrated in Fig. \ref{FIG:RWLD}.

\subsection{Large deviations with additional constraints}
\label{sec:addconst}

We have argued that large deviations of the mean
\[ \frac{1}{N}\sum_{i=1}^N X_i =\bar{x}>\langle X\rangle_Q \]
in (right) fat tailed distributions are characterized by concentrated realizations of the sample $\underline{X}=\{X_1,\ldots, X_N\}$.
In this case, the emergence of an extensive variables $X_{i^*}\simeq N(\bar{x}-\langle X\rangle_Q)$ leads to the divergence of all sample moments $\frac{1}{N}\sum_{i=1}^N X_i^k$ with $k>1$, as $N\to \infty$. Indeed, at the leading order:
\[ \frac{1}{N}\sum_{i=1}^N X_i^k = \frac{1}{N}\sum_{i\neq i^*} X_i^k + \frac{1}{N}X_{i^*}^k \simeq N^{k-1}(\bar{x}- \langle X\rangle_Q)^k \ , \]
which diverges for $N\to\infty$. In this section we investigate the behavior of the sample $\underline{X}$ in the case where the mean is deviating but the $k$-th sample moment is fixed to a finite value. Since a finite value of the moment is not typical, this corresponds to investigating simultaneous large deviations of both the mean and the $k$-th moment of the sample.
Hence, let us consider the following events:
\[ E=\left\{ \frac{1}{N}\sum_{i=1}^N X_i \in [\bar x,\bar x+\delta x]\right\} \]
and
\[ E'=\left\{ \frac{1}{N}\sum_{i=1}^N X_i^k \in [m_k,m_k+\delta m]\right\} \qquad (\mathrm{for}\; k>1) \ , \]
which denote large deviations of the mean and the $k$-th moment, respectively. Our purpose is to study the joint probability $P\{E,E'\}$, denoting the probability that both events $E$ and $E'$ occur at the same time, in order to understand if concentration phenomena take place. For infinitesimal $\delta x$ and $\delta m$, the probability $P\{E,E'\}$ can be written in terms of the joint p.d.f.\ $\tilde\rho_N(\bar{x},m_k)$, namely, $P\{E,E'\} = \tilde\rho_N(\bar{x},m_k)\delta x\delta m$. For a fat tailed distribution $Q(x)$ with positive support, $\rho_N(\bar{x},m_k)$ can be expressed as a multiple integral with two constraints on the mean (event $E$) and the moment (event $E'$), respectively:
\[ \tilde\rho_N(\bar{x},m_k) = \int \left(\prod_{i=1}^N \upd x_i\,q(x_i) \right) \df{\frac{1}{N}\sum_{i=1}^N x_i - \bar{x}} \df{\frac{1}{N}\sum_{i=1}^N x_i^k - m_k} \ . \]
In the following, we assume again that the distribution $Q(x)$ has finite mean and variance, such that both the Law of Large Numbers and the Central Limit Theorem hold. We will consider both the case where 
the expected value $\langle X^k\rangle_Q$ is finite, i.e. when $q(x)$ decays at least as fast as $|x|^{-k-1}$, and when $\langle X^k\rangle_Q=+\infty$.

As in Sec.\ \ref{sec:thintails}, the integral expression for $\tilde\rho_N(\bar{x},m_k)$ can be evaluated by means of a saddle point approximation within the density functional formalism. Indeed, in the absence of concentration, we can rewrite it in terms of the density $p(x)=\frac{1}{N}\sum_{i=1}^N\df{X_i-x}$, thus getting:
\begin{eqnarray*}
\tilde\rho_N(\bar{x},m_k) &\approx& \int \mathcal{D}p(x)\, e^{-ND_{KL}(P||Q)} \df{\int\upd x\, p(x) - 1}\times \\
&\times& \df{\int\upd x\,x\,p(x) - \bar{x}} \df{\int\upd x\,x^k\,p(x) - m_k} \ .
\end{eqnarray*}
The constrained minimization of $D_{KL}(P||Q)$ leads to the saddle point solution:
\begin{equation}
p^*(x)=p_{\beta,\nu}(x)=\frac{1}{Z(\beta,\nu)}q(x)e^{-\beta x -\nu x^k} \ ,
\label{SP_SOLUTION}
\end{equation}
where $Z(\beta,\nu)$ is a normalization factor and $\beta$ and $\nu$ are adjusted in order to obtain:
\begin{eqnarray}
\label{SP_CONDITIONS}
\langle X\rangle_{P_{\beta,\nu}} &=& \frac{1}{Z(\beta,\nu)}\int\upd x\, x\,q(x)\,e^{-\beta x -\nu x^k} = \bar{x} \\ \nonumber
\langle X^k\rangle_{P_{\beta,\nu}} &=& \frac{1}{Z(\beta,\nu)}\int\upd x\, x^k\,q(x)\,e^{-\beta x -\nu x^k} = m_k \ .
\end{eqnarray}
The result (\ref{SP_SOLUTION}) holds as long as Eqs.\ (\ref{SP_CONDITIONS}) have a solution in $\beta$ and $\nu$. 

\begin{figure}
\centering
\includegraphics[width=0.45\textwidth]{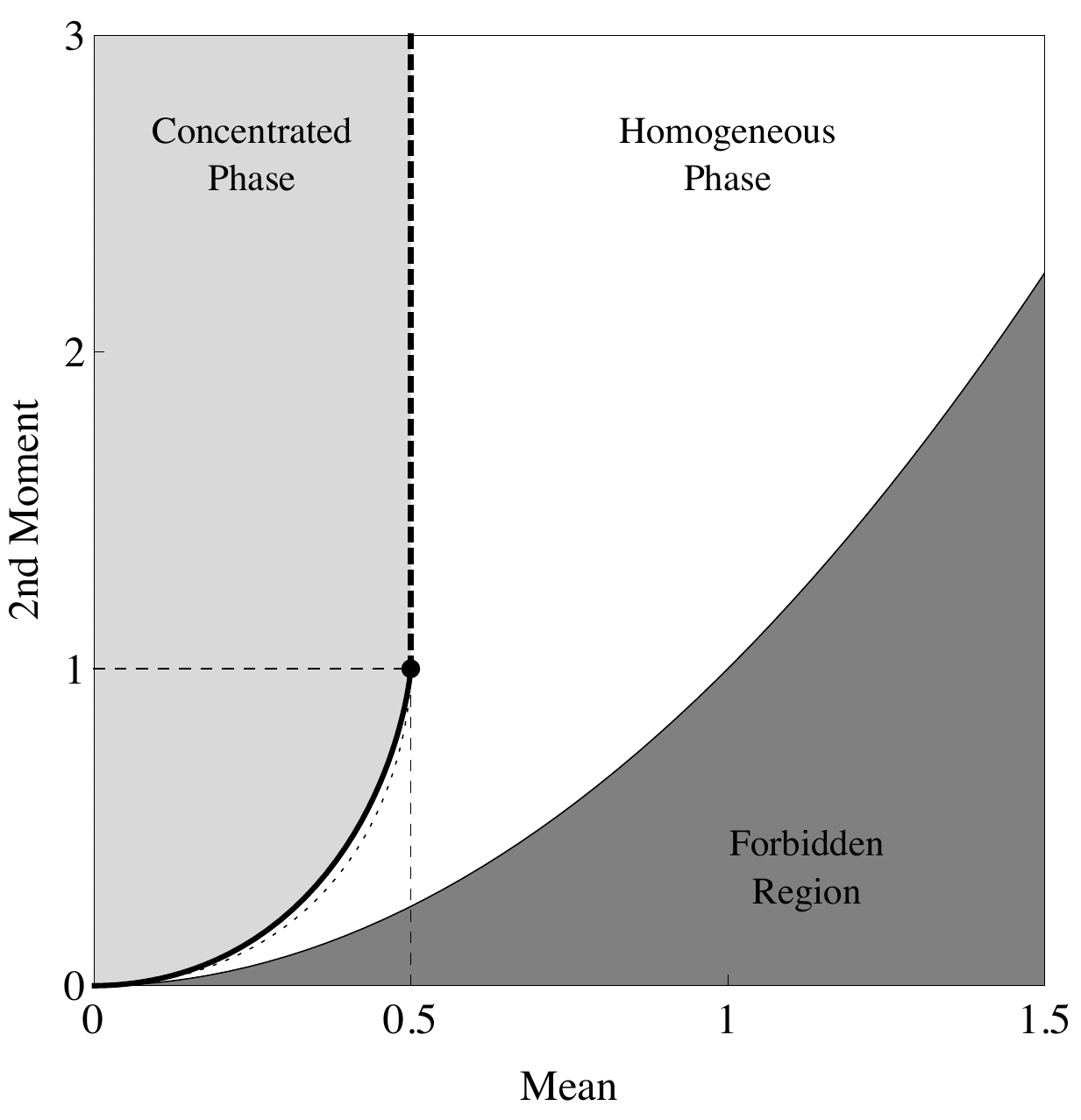}
\hfill
\includegraphics[width=0.45\textwidth]{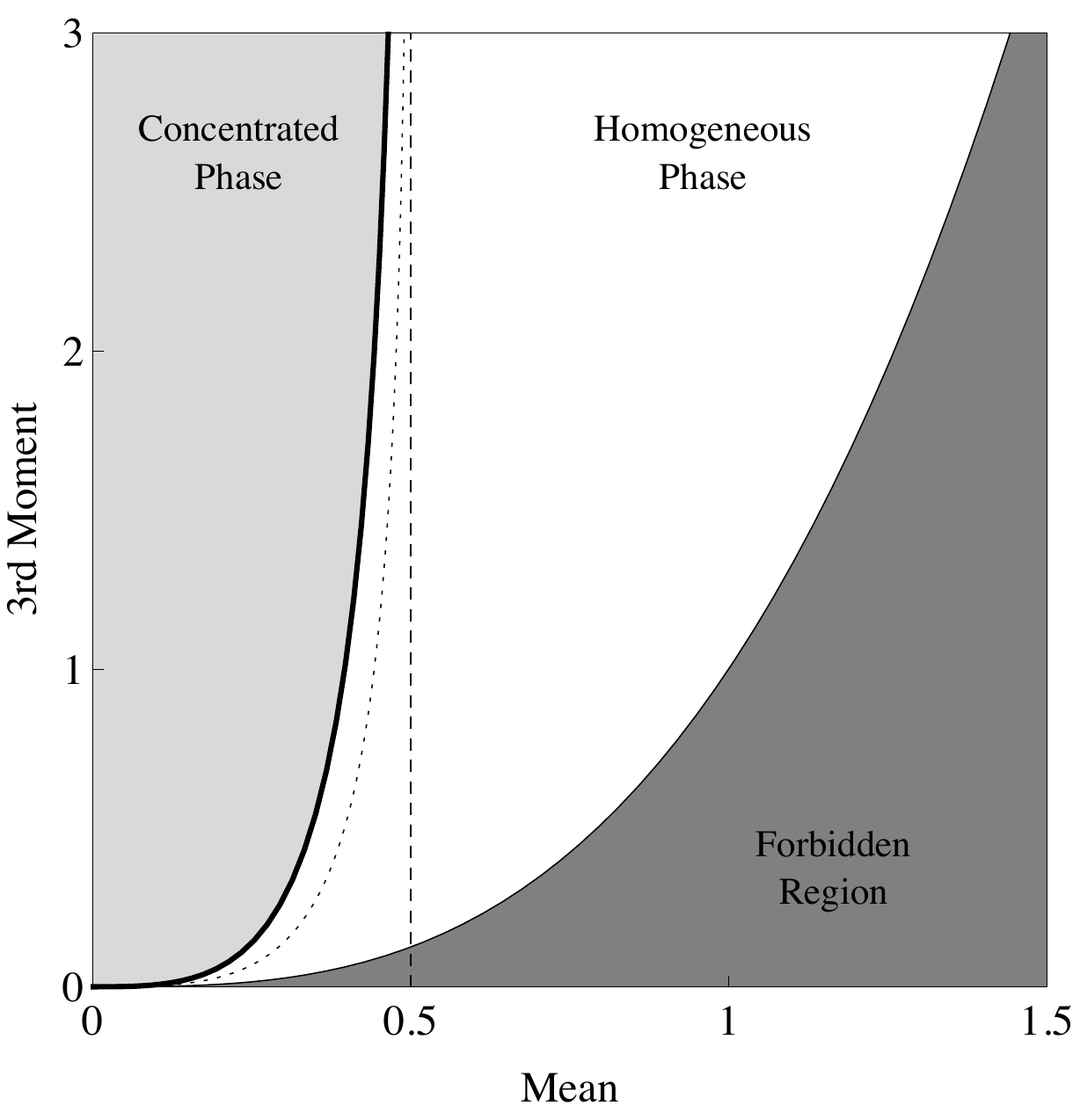}
\caption{Phase diagrams of large deviations of i.i.d.\ random variables at fixed mean $\bar x$ and moment $m_k$ for a fat-tailed distributions $Q(x)$. Two cases are shown, where $k$ has been choosen such that $\langle X^k\rangle_Q$ either converges (left) or diverge (right). The straight dashed lines denote the critical values $\bar x=\langle X\rangle_Q$ and $m_k=\langle X^k\rangle_Q$. The bold line (i.e.\ the phase boundary) and the dotted line are obtained for $\nu=0$ and $\beta=0$, respectively (see Eq.\ (\ref{SP_CONDITIONS})). The dark region is forbidden by the Jensen inequality: $m_k \geq (\bar{x})^k$. No concentration can be observed for $\bar x >\langle X\rangle_Q$. The plot has been obtained with a shifted Pareto distribution $p(x)=\alpha (x+1)^{-(\alpha+1)}$ with $x\geq0$ and $\alpha=3$, choosing $k=2$ (left) and $k=3$ (right).}
\label{fig_phdiag3}
\end{figure}

Fig. \ref{fig_phdiag3} illustrates the behavior of the solution for a representative choice of $Q(x)$ where $\langle X^k\rangle_Q$ is finite (left) or not (right). The case where the mean of $\langle X^k\rangle_Q$ attains its typical value, which has been discussed in previous sections, is recovered when $\nu=0$. This corresponds to the bold line that separates the light shaded and the white region. When $\langle X^k\rangle_Q=+\infty$ (Fig. \ref{fig_phdiag3} right) the line extends to infinity on the vertical direction, whereas if $\langle X^k\rangle_Q<+\infty$ (Fig. \ref{fig_phdiag3} left) this line ends at the point where both the first and the $k^{\rm th}$ moment take their typical value. When the first moment is forced to attain a non-typical value $\bar x> \langle X\rangle_Q$, as discussed above, the $k^{\rm th}$ moment diverges. So the trajectory of samples in the $\bar x$--$m_k$ plane as $\bar x$ varies is described by the bold line for $\bar x\le \langle X\rangle_Q$ and by an horizontal line at infinity for $\bar x> \langle X\rangle_Q$. For $\langle X^k\rangle_Q<+\infty$, this $\nu=0$ trajectory has a vertical jump (dashed line in Fig. \ref{fig_phdiag3} left). 

The line $\nu=0$ separates two regions, that have statistically different properties. On the right of the $\nu=0$ line (white region) we find a solution with $\nu>0$ and $\beta$ that is adjusted to enforce $ \langle X\rangle_Q = \bar x$. In particular, $\beta=0$ is achieved on the dotted line and $\beta>0$ to the left of the dotted line. 
Remarkably, at odds with the case discussed in the previous section, no concentration arises for $\bar x> \langle X\rangle_Q$ as soon as a constraint on a higher moment is introduced. 

The dark shaded region to the right of the white region has $m_k<\bar x^k$. By Jensen's inequality, $\langle |X|^k\rangle_P \ge \langle X\rangle_P$ for any distribution $P$. Therefore no point in this region can be achieved. The points on the boundary correspond to distributions $P$ with a point mass at $\bar x$ and are realized in the limit
$\nu\to \infty$ with $\beta$ also diverging in a specific combination.\ \footnote{For $k=2$ it is easy to check that points with $\langle X^2\rangle_P =\bar x^2$ can be achieved in the limit $\nu\to\infty$ with $\beta/\nu=2\bar x$.}

Finally, when $\bar x$ and $m_k$ attain values in the region to the left of the $\nu=0$ line (grey region in Fig. \ref{fig_phdiag3}) the solution derived above breaks down and we must apply, instead, a \emph{concentration ansatz}. As in Sec.\ \ref{sec:funcint_cond}, the ansatz consists in breaking the system's symmetry under exchange of  variables. This can be achieved by excluding one variable, say $X_{i^*}$, from the definition of $p(x)$ and rewriting it in terms of the remaining $N-1$ variables:
\begin{eqnarray*}
&&\tilde\rho_N(\bar{x},m_k) \approx \int \mathcal{D}p(x)\,\upd x_{i^*} \, e^{-F(P,Q)} \df{\int\upd x\, p(x) - 1} \times \\
&\times& \df{\frac{N-1}{N}\int\upd x\,x\,p(x) +\frac{x_{i^*}}{N} - \bar{x}} \df{\frac{N-1}{N}\int\upd x\,x^k\,p(x) +\frac{x_{i^*}^k}{N} - m_k} \ ,
\end{eqnarray*}
where  $F(P,Q) = (N-1) D_{KL}(P||Q) - \ln q(x_{i^*})$. In order to solve the integral with the saddle point approximation, we again rescale $p(x)$ and $x_{i^*}$ such that all the leading terms have the same scaling in $N$. Assuming that $k>1$, the only non-trivial scaling is given by $x_{i^*}\sim N^{1/k}$. Thus, we perform the substitutions $x_{i^*}=N^{1/k}t$ and, at the leading order in $N$, we obtain:
\begin{eqnarray*}
\tilde\rho_N(\bar{x},m_k) &\approx& \int \mathcal{D}p(x)\,\upd t \, e^{-ND_{KL}(P||Q)} \df{\int\upd x\, p(x) - 1} \times \\
&\times& \df{\int\upd x\,x\,p(x) - \bar{x}} \df{\int\upd x\,x^k\,p(x) + t^k - m_k} \ .
\end{eqnarray*}
Finally, the constrained minimization of $D_{KL}(P||Q)$ leads to the following saddle-point solution:
\begin{equation}
p^*(x)=p_\beta(x)\equiv \frac{1}{Z(\beta)} q(x) e^{-\beta x} \qquad\mathrm{and}\qquad t^* = \bigl(m_k-\langle X^k\rangle_{P_\beta}\bigr)^\frac{1}{k} \ ,
\label{SP_SOLUTION_CONC}
\end{equation}
where $\beta$ is such that $\langle X \rangle_{P_\beta}=\bar x$. We remark that 
since $\bar x\leq\langle X \rangle_Q$ in this region, the normalization integral $Z(\beta)$ is finite because $\beta\geq0$.

We conclude that, also in this case, the large deviation concentrates in the single variable $X_{i^*}$, which scales as $X_{i^*}\simeq N^{1/k} (m_k-\langle X^k\rangle_{P_\beta})^{1/k}$. Unlike the case where $\bar x$ is fixed but $m_k$ is not constrained -- where $X_{i^*}\simeq N(\bar{x}-\langle X\rangle_Q)$ -- here the concentrating variable is sub-extensive ($k>1$) and carries a finite fraction of the moment $m_k$ rather than of the mean $\bar x$. The remaining variables, instead, are not influenced by the constraint on $m_k$ but are affected by the constraint on $\bar x$, which modifies their marginal distribution by a damping factor $e^{-\beta x}$. Loosely speaking, the excess moment $m_k-\langle X^k\rangle_Q$ concentrates on just one variable, whereas the excess mean $\bar x-\langle X\rangle_Q$ is equally distributed among all other variables.

The two different realizations of the combined large deviations of $E$ and $E'$, namely, the homogeneous solution (\ref{SP_SOLUTION}), and the concentrated solution (\ref{SP_SOLUTION_CONC}), can also be described as a phase transition. When the saddle point solution $p^*(x)$ passes from $p_{\beta,\nu}(x)$ to $p_\beta(x)$, the rate function $I(\bar{x},m_k)=D_{KL}(P_{\beta,\nu}||Q)$ becomes non-analytic. The fact that concentration of  large deviations in the mean disappears as soon as a constraint on an higher moment is introduced, as well as the fact that the higher moment concentrates only if the mean is less than the expected value, are apparently paradoxical results. 

As a final remark, we recall that the marginal distribution (\ref{SP_SOLUTION}) in the homogeneous phase is modified by the additional damping factor $e^{-\nu x^k}$, which could lead to a divergent distribution for $\nu<0$. Therefore, the concentration of large deviations does not occur only in fat tailed distibutions, but can be observed also in other distributions with thinner tails. More specifically, large deviations of the $k$-th moments (above the critical point) concentrates in any distribution $Q(x)$ such that $e^{hx^k} q(x)$ diverges for all $h > 0$, when $x\to\infty$. For example, large deviations of the 3rd moment are concentrated even in random samples with normal distribution.\ \footnote{This relates to the observation in \cite{Cov} that maximum entropy distributions may be realized in anomalous manner. E.g. the maximum entropy distribution when the first three moments are fixed may be realized by a Gaussian that enforces the first two moments, with a vanishingly small mass $\epsilon$ at a point $x_0= c\epsilon^{1/3}$ where $c$ is fixed so as to enforce the constraint on the third moment, in the limit $\epsilon\to 0$ (see \cite{Cov} section 12.3).}
Although they have been derived through a different procedure, the above results agrees with the recent analysis performed by Szavits-Nossan, Evans and Majumdar \cite{Szavits-Nossan}.

\section{Large deviations for non independent variables}
\label{sec:ldrmt}

Clearly the results of the previous section do not generalize to cases with non independent variables. For example, if $X_1,\ldots,X_n$ have a joint elliptic distribution
\[ p(x_1,\ldots,x_n)=\int_0^\infty\!d\sigma \rho(\sigma)e^{-\frac{1}{2\sigma^2}\sum_i x_i^2} \ , \]
each variable $X_i$ can have a fat tailed marginal distribution $p(x)$ provided the distribution $\rho(\sigma)$ of $\sigma$ is fat tailed. Yet it is easy to see that large deviations do not concentrate. Indeed, conditional on $\sigma$ the variables $X_1,\ldots, X_n$ are Gaussian and one can reduce a large deviation $\sum_i X_i=N\bar x$ to the occurrence of an anomalously large value of $\sigma$. Typical values of a sum of $N$ Gaussian variables is of order $\sqrt{N}$, therefore $\sum_i X_i\sim\sigma\sqrt{N}$. A large deviation $\sigma\sim \sqrt{N}$ is enough to achieve a large deviation in the sum. 

\subsection{Eigenvalues of random matrices}
\label{sec:EIGEN_RM}

The eigenvalues of a random matrix with entries independently drawn from a given probability distribution are, in general, non-independent random variables. One of the simplest realizations of this phenomenon, originally considered by Wigner \cite{Wigner}, is that of $N \times N$ real symmetric matrices with elements $M_{i,j}$ drawn at random from a Gaussian distribution with zero mean and variance scaling as $N^{-1}$. Another case is represented by the Wishart ensemble, originally introduced in \cite{Wishart}, where
\begin{equation} \label{cov}
M_{i,j}=\frac{1}{T}\sum_{t=1}^T x_{i,t} x_{j,t} \ ,
\end{equation}
and the $x_{i,t}$'s are i.i.d. random variables with zero mean and finite (typically unit) variance. Here $T=q N$ is taken proportional to $N$. Such an ensemble can often be treated as a ``zero order'' null model for correlation analysis in multivariate statistics, see e.g. \cite{Johnstone}. As such, it has been employed in an impressive list of applications ranging from wireless communications \cite{Tulino} and image processing \cite{Fukunaga} to financial data analysis \cite{BouchaudPotters}, which later will be later addressed in this paper.

Under mild hypotheses on the probability distribution $Q(\hat{M})$ of the random matrix ensemble, the eigenvalues $\lambda_1\ge\lambda_2\ge\ldots\ge\lambda_N$ of the matrix $\hat M$ exhibit typical properties in the sense that their density $\rho(\lambda)d\lambda$, i.e. the fraction of $\lambda_i\in [\lambda,\lambda+d\lambda)$, attains a non trivial limit when $N\to\infty$. The two aforementioned cases are good examples of this: the average eigenvalue density of random symmetric matrices approaches, in the large $N$ limit, the celebrated Wigner's semicircle distribution \cite{EdwardsJones}, whereas the corresponding density for matrices of the type in equation (\ref{cov}) approaches the Mar\v cenko-Pastur distribution when the thermodynamic limit $N,T\to\infty$ (with $q=T/N$ kept fixed) is taken (see e.g. \cite{BurdaJurkiewicz}). In both such cases, the eigenvalue density has support on a finite interval $[\lambda_-,\lambda_+]$ and therefore the average of the eigenvalues
\[
\bar\lambda=\frac{1}{N}\sum_{i=1}^N \lambda_i
\]
converges almost surely to the expected value
\[
\langle \lambda \rangle_Q =\int\! d\lambda \rho(\lambda)\lambda \ .
\]

We consider large deviations $E=\left\{\sum_i \lambda_i=N\bar\lambda\right\}$ where $\bar\lambda > \langle \lambda \rangle_Q$ takes anomalously large values.\ \footnote{We do not discuss large deviations with $\bar\lambda< \langle \lambda \rangle_Q$ that are much more rare ($\sim e^{-aN^2}$) in the Wishart ensemble \cite{satya3}.}
When the underlying distribution that defines $Q(\hat M)$ has thin tails (e.g. Gaussian), the probability of the event $E$ is exponentially small in $N$. In the Wigner case the large deviation is realized by a shift of the eigenvalue distribution, that corresponds to adding a term proportional to the identity to $\hat M$, i.e: $\hat M\to \hat M+b\hat I$ with $b=\bar\lambda-\langle \lambda \rangle_Q$. In the Wishart ensemble with thin tails, the large deviation is realized by dilatation $\lambda_i\to b\lambda_i$, i.e. $\rho(\lambda)\to \rho'(\lambda)=\rho(\lambda/b)/b$ with $b=\bar\lambda/\langle \lambda \rangle_Q$ \cite{Nadal}. More sophisticated large deviations investigations of the Wishart ensemble have been recently performed in \cite{Biroli}, and in \cite{Vivo,satya2,Vivo2} by means of statistical mechanical techniques consisting of mapping the eigenvalues onto a one dimensional gas of interacting particles.

Here we show that, when the underlying distribution is fat tailed, in both cases the large deviation is realized by concentrating the large deviation on the largest eigenvalue, that corresponds to adding a rank one matrix to $\hat M$, i.e: $\hat M\to \hat M+\ket{v}\bra{v}$.\ \footnote{We adopt the bracket notation: column vectors are denoted as $\ket{v}$ and row vectors as $\bra{v}$. $\braket{v}{u}$ is the scalar product whereas $\ket{v}\bra{u}$ is the tensor product.}

The key observation leading to this result is that the average of the eigenvalues is simply related to the trace of $\hat M$ and that has in both cases a simple expression in terms of the underlying random variables
\begin{eqnarray}
\bar\lambda & = & \frac{1}{N}\sum_{i=1}^N M_{i,i} \qquad \hbox{Wigner} \\
 & = & \frac{1}{NT}\sum_{t=1}^T\sum_{i=1}^N x_{i,t}^2 \qquad \hbox{Wishart} \ .
\end{eqnarray}
Therefore the large deviation in the eigenvalues corresponds to a large deviation in the sum of i.i.d. variables ($M_{i,i}$ and $x_{i,t}^2$ respectively for the two cases). Under a fat tailed distribution we expect that this sum will be dominated by one large element, carrying the whole excess deviation. 

Note in particular that, in the Wishart case, the concentration of large deviations in the sum of $x_{i,t}^2$ requires the distribution of $x_{i,t}^2$ to be fat tailed, i.e. to fall off slower than an exponential at infinity.
This implies that large deviations concentrate whenever the p.d.f. $q(x)$ of $x_{i,t}$ dacays slower than a Gaussian for $|x|\to\infty$.
Then a large deviation with $\bar\lambda>\langle \lambda \rangle_Q$ will typically be realized by samples where for some indices ${i^*,t^*}$ one has $x_{i^*,t^*}^2\simeq NT\left(\bar\lambda-\langle \lambda \rangle_Q\right)$.
Notice that this requires $x_{i^*,t^*}\sim \sqrt{NT}$ which is proportional to $N$ when $T\sim N$. Likewise, in the Wigner case, the large deviation will ``localize'' on a single anomalously large value $M_{i^*,i^*}\simeq N(\bar\lambda-\langle \lambda \rangle_Q)$ along the diagonal.

Without loss of generality we can take $i^*=1$, so that the corresponding random matrix has the form
\[ M_{k,j}^{(LD)}=M_{k,j}^{\rm typ}+N\left(\bar\lambda-\langle \lambda \rangle_Q\right)\delta_{k,1}\delta_{j,1}=
\left(\begin{array}{cc}N \delta\lambda & \langle b | \\ | b \rangle & \hat C\end{array}\right) \ , \]
where $\delta \lambda = \bar\lambda-\langle \lambda \rangle_Q$. To a first approximation, this has the form of a two-blocks matrix 
\[ \hat M^{(0)}= \left(\begin{array}{cc}N\delta\lambda & 0 \\ 0 & \hat C\end{array}\right), \]
where the first row/column refers to index $i^*$, and the second to the other $N-1$ indices.
The eigenvalue spectrum of $\hat M^{(0)}$ is exactly the same of the original ensemble with an additional eigenvalue $\lambda_1=N\delta\lambda$ that is well separated. The eigenvalue distribution of the original matrix $\hat M^{(LD)}$ is worked out within perturbation theory in \ref{app:perturbation_theory}, considering 
\[ \hat V=\left(\begin{array}{cc}0 & \bra{b} \\\ket{b} & 0\end{array}\right) \]
as a perturbation. The result confirms that the correction to the eigenvalue distribution of $\hat M^{(0)}$ is indeed small. This proves that large (positive) deviations in the sum of eigenvalues of random matrices derived from an underlying fat tailed distribution concentrate on the largest eigenvalue. In addition, the discussion above also shows that the corresponding eigenvector is localized on a single index $i^*$ (and on a single ``time'' $t^*$ for Wishart matrices).

We would like to conclude this section with some remarks.
The eigenvalues of Wigner and Wishart matrices are a special case of dependent variables subject to a ``repulsive'' interaction, specifically, the same interaction of Coulomb charges in two dimensions. In this case, the sum of interacting variables can be reduced to the sum of i.i.d.\ random variables thanks to the rotational invariance of the trace. Yet, there are similar cases where this invariance does not hold and the interaction may give raise to interesting and uncommon phenomena.
An example is given by the heights of non-intersecting interfaces on a substrate \cite{Nadal_PRE}, whose joint distribution can be mapped to the joint distribution of Wishart's eigenvalues. Indeed, this system exhibits an {\em infinite order phase transition}, occurring when the sum of the height crosses its expected value (hence, at the same critical point of the concentration phase transition). These variables are thin-tailed variables and do not concentrate: it could be interesting to investigate a possible fat-tailed version of this system, where an interplay between the concentration transition and the infinite-order transition is expected.

As a final comment, we point out that the concentration of large deviations in the eigenvalues of random matrices can be obtained not only through the constraint $\sum_i\lambda_i=N\bar\lambda$, but also through an additional constraint on a higher moment, say, $\sum_i\lambda^k=N\lambda^{(k)}$ with $k>1$, even though there is no direct relationship between the moment and the trace of the matrix. This is in line with the result of Sec.\ \ref{sec:addconst} concerning i.i.d.\ random variables. This issue has been discussed in great details in \cite{Nadal} and \cite{Nadal_PRL} in the context of quantum entanglement of bipartite states, to which we refer the interested reader.

\section{Application to financial data}
\label{sec:findata}

Finance offers several instances where the results we have discussed are relevant. 
Several risk measures used in risk management of large portfolios \cite{BouchaudPotters} are based on conditional losses in the tails. This is a clear examples of a large deviation, because the loss of the portfolio is the sum of the individual losses on the assets.  
When the p.d.f.\ of individual losses is fat tailed, which is the case for equities \cite{Stanley}, then our discussion suggests that typical losses will be realized in samples where one of the assets  drops by a much larger amount than the others. For example, 
in a portfolio of $N=100$ well diversified stocks with a p.d.f.\ of returns that decays as $|x|^{-4}$ \cite{Stanley}, the
Value-at-Risk (see e.g. \cite{Jorion}) at 1\% is expected to be dominated by the largest drop, that should carry more than 56\% of the total loss.\ \footnote{Indeed returns in a financial market are correlated. Applying this analysis to daily returns of $N=41$ stocks in the Dow Jones index from the period 1980-2005 (see \cite{RaffaelliMarsiliPonsot} for a description of the data) we found that in the worst portfolio return in a year (i.e. over 250 days, i.e. 0.4\%) losses concentrate for 12\% on a single stock. This is much less than what one would get with i.i.d.\ variables with a distribution $p(x)\sim |x|^{-4}$ (72\%), but it is still significant compared to the case of $N=41$ Gaussian i.i.d.\ variables where the largest loss accounts only for 2.6\% of the total. A higher concentration is observed for the best returns in a year, where the stock with the highest return carries 21\% of the weight. This reflects the fact that correlations in the negative tail of the joint p.d.f.\ of returns are stronger than in the positive tail.}

The problem of estimating credit risk is again of the same nature, as one focuses on events where the equity of a company, which is a sum on the different lines of business, becomes negative. Again if returns from the different investments are broadly distributed, and within the simplest approximation where they are considered independent, we expect default events to be characterized by a similar ``pernicious concentration of bad luck''. We refer to Ref. \cite{Embrechts} for a more detailed discussion of risk management. 

The rest of this section deals with relating the insights discussed in the previous section to the statistics of price fluctuations of financial stocks. 

\subsection{Large excursions of prices and jumps}

\begin{figure}
\begin{center}
\includegraphics[width=0.6\textwidth]{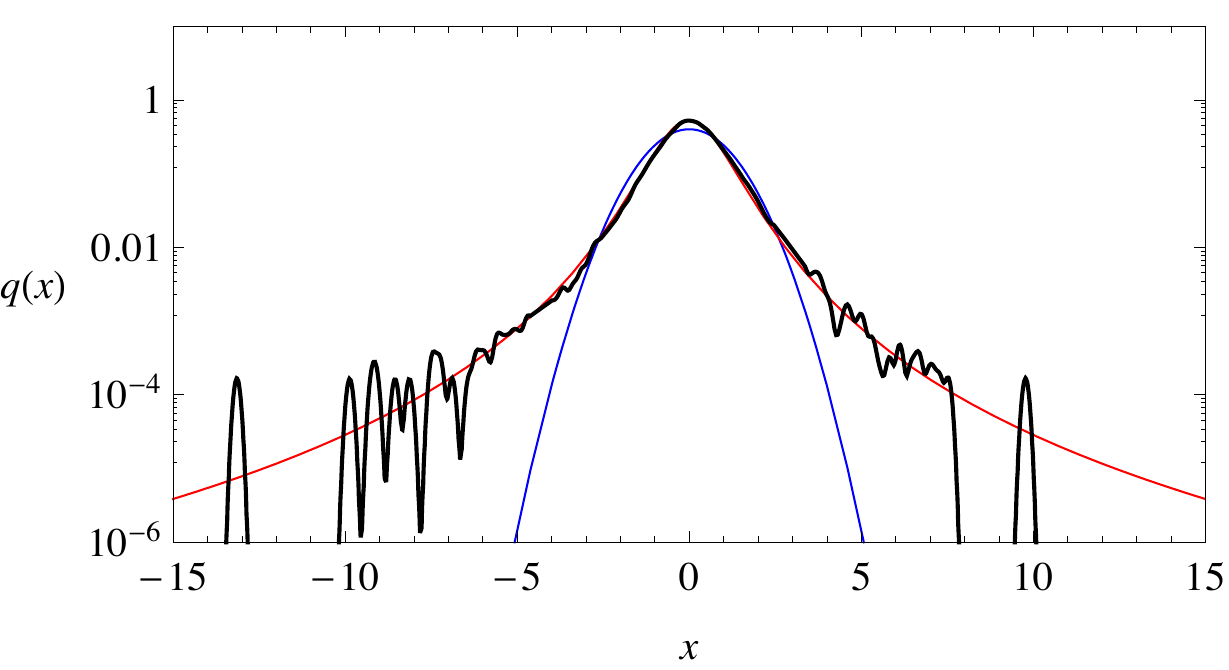}
\end{center}
\caption{The empirical distribution of price returns (black) from dataset D1 (see Table \ref{TABLE}), compared with the normal distribution (blue) and the Student's t-distribution with 4 degrees of freedom (red).}
\label{FIG:PDF}
\end{figure}

The price $P(t)$ of a stock at time $t$ can be considered as a geometric stochastic process
\[ \log P(t_N) = \log P(t_0) + \sum_{i=1}^N r(t_i) \]
driven by the logarithmic return
\[ r(t_i) = \log \frac{P(t_i)}{P(t_{i-1})} \ . \]
To a zeroth order approximation, we can assume that:
\begin{description}
\item[A] the distribution of returns is fat-tailed;
\item[B] the returns are i.i.d.\ random variables.
\end{description}
The assumption {\bf A} reproduces an empirical fact. It is well-known (see \cite{Stanley,BouchaudPotters,Cont}, for instance) that the marginal distribution of returns is a power-law distribution with an exponent in the range $3 \div 4$, and this feature is also confirmed by financial data at our disposal (see Fig.\ \ref{FIG:PDF}).
The assumption {\bf B}, instead, is the simplest translation of the so-called Efficient Market Hypothesis, which states that financial markets are informationally efficient \cite{Fama}. The statement {\bf B} is not verified by empirical observations: for example, even though returns are not linearly auto-correlated, their absolute values exhibit a strong auto-correlation in time with long-memory effect \cite{Cont,Taylor}. 
Yet, it makes sense to compare the behavior of large deviations of returns in real data with the prediction of the theory discussed above. This provides clear indication of how hypothesis {\bf B} is violated and what precisely the role of statistical dependencies is as far as large deviations are concerned.

As we already argued (see Sec.\ \ref{sec:stocproc}, for instance), the assumptions {\bf A} and {\bf B} imply that large excursions of stock prices are concentrated in isolated returns, namely, that they are more likely generated by large discontinuous jumps rather than by continuous drifts. Discontinuous movements of stock prices 
can be actually detected in financial time series and have attracted great attention in recent financial literature \cite{Kirilenko,Bormetti}.
In the following analysis, we will compare the theoretical expectations based on the assumption {\bf A} and {\bf B} with some empirical observation on real financial data, in order to check if the occurrence of concentration phenomena in financial time series is compatible with our hypothesis.

\begin{figure}
\centering
\includegraphics[width=\textwidth]{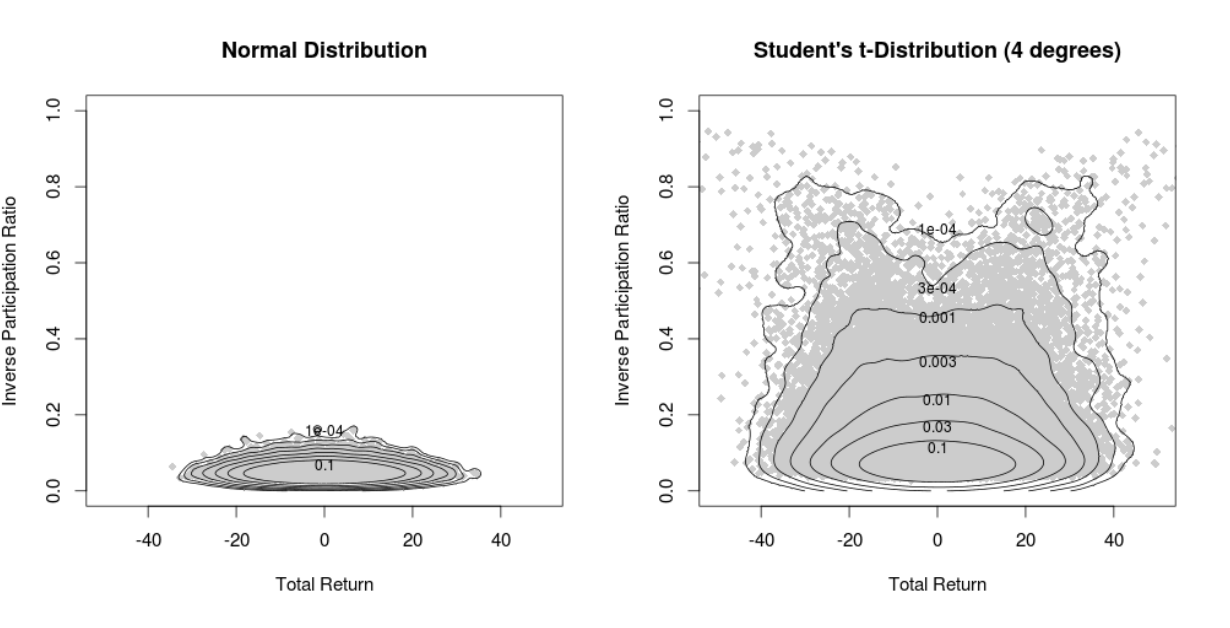}
\caption{Scatter-plots of $I_2$ vs $R$ for i.i.d. random returns drawn from a standard normal distribution (left) and from a standard Student's $t$-distribution with 4 degrees of freedom (right). The plots are obtained from 213,000 observations of $N=60$ returns,
corresponding to the same number of observations performed in Fig.\ \ref{FIG:MAINPLOT} on the experimental dataset D2 (see Table \ref{TABLE}).
The different behavior of the IPR clearly shows the absence/presence of concentration phenomena in the normal/Student's case, respectively.
The iso-probability contours of the joint probability density of $R$ and $I_2$ are shown (the density is evaluated through a Gaussian kernel estimation with bandwiths $(1.0,0.01)$ and $(2.0,0.02)$ for the normal and Student's case, respectively).}
\label{FIG:SIMULATION}
\end{figure}

In order to analyze how large excursions of stock prices are realized, we measure price returns at two different time-scales $\Delta t$ and $\delta t$, with $\Delta t \gg \delta t$.
The return $R$ over the period $\Delta t$ is given by the sum of many consecutive returns $r_1, r_2, \dots, r_N$ over the smaller time steps $\delta t$, namely:
\[ R = \sum_{i=1}^N r_i \ , \]
where $N = \Delta t/\delta t$.
The degree of concentration of the return $R$ in term of its components $r_1,r_2,\dots,r_N$ can be measured by means of the generalized Inverse Participation Ratio (IPR)
\[ I_k = \left(\sum_{i=1}^{N} w_i^k\right)^\frac{1}{k-1} \ , \]
where
\[ w_i = \frac{r_i^2}{\sum_{j=1}^N r_j^2} \]
is the squared weight of the $i$-th return with respect to the \emph{sample volatility}, i.e.\ the typical scale of fluctuations of price returns in the sample $r_1,r_2,\dots,r_N$.
The weights $w_i$ define a probability measure (they are always between 0 and 1 and sum up to 1), therefore $I_k$ is always in the range $[1/N,1]$. If the deviation of $R$ from its expected value is equally distributed among all variables $r_i$, then all weights $w_i$ are equal to $1/N$. In this case the IPR takes its minimum value ($I_k=1/N$) and vanishes in the limit $N\to\infty$. On the other hand, if the deviation of $R$ is concentrated in the single variable $r_{i^*}$, one gets $w_{i^*}=1$ with all other weights vanishing. In this case, the IPR takes its maximum value ($I_k=1$) and remains finite for $N\to\infty$.
We focus our analysis on the specific choice $k=2$ (the classical value of the IPR).
It is worth noticing that $I_k=\exp(-S_k[w_i])$, where $S_k[w_i]$ is the Renyi entropy of the weights $w_i$, another well-known measure of concentration/dispersion in a probability distribution.

In Fig.\ \ref{FIG:SIMULATION} we show the relative behavior of $I_2$ and $R$ for i.i.d.\ random returns $r_1, r_2, \dots, r_N$ drawn from a normal distribution and from a Student's $t$-distribution with 4 degrees of freedom. The figure highlights the difference between concentrated and non-concentrated realization of the sum of random variables, showing how the IPR is a good measure of concentration phenomena.

In order to test our expectation, we observe the time series of many stock prices and we repeat several measurements of the price return $R$ over the time-scale $\Delta t$. For each return, we measure its components $r_i$ over the finer time-scale $\delta t$ and we evaluate their IPR $I_k$. Then, as in Fig.\ \ref{FIG:SIMULATION}, we analyze the dependence of $I_k$ with respect to $R$ by means of a scatter-plot, checking if large values of $R$ are actually associated with large values of $I_k$ (and vice-versa).

\begin{table}
\footnotesize
\centering
\begin{tabular}{p{.20\textwidth}p{.36\textwidth}p{.36\textwidth}}
\hline
& \textbf{First Dataset (D1)} & \textbf{Second Dataset (D2)} \\
\hline
\bf Market & NYSE 100 & FTSE 40 \\
\bf & U.S. Stock Market & Italian Stock Market \\
\hline
\bf Period & Feb 2001 - Dec 2003 & Apr 2012 - Aug 2013 \\
& 727 working days & 355 working days \\
\hline
\bf Stocks & 100 most traded stocks & 40 most traded stocks \\
\hline
\bf Type of data & All trades: execution time, trading price, and exchanged quantity. & Price and quantity of the best quotes (best-ask and best-bid), updated every 1 sec. \\
\hline
\bf Measured price & Last-execution price & Mid-price (mean value between the ask and bid prices) \\
\hline
\end{tabular}
\caption{Details about the datasets used for the analysis of concentration phenomena in financial time series.
Data have been kindly provided by M.I.U.R.\ (D1) and by LIST S.p.A.\ (D2).}
\label{TABLE}
\end{table}

We analyzed two different datasets (D1 and D2) of financial time series, whose main features are described in Table \ref{TABLE}. They are very different datasets: they refer to different markets and periods, and they have different length and time-resolution. Such heterogeneity allows us to check the stability of our results across different periods, markets, and time-scales. The time series in D2 are shorter than in D1, but they are based on quotes updates rather than on trades executions, so they allow a direct measure of the mid-price. At high frequencies, the mid-price is usually a better measure than the last-execution price: it is constantly updated an it is less affected by market micro-structure effects (e.g.\ the bid-ask bounces).
For these reasons, we decided to exploit D1 for the analysis over daily time-scales, and D2 for the analysis at smaller time-scales (from few minutes to one day). In the choice of the time intervals $\Delta t$ and $\delta t$ we have to regard the following criteria:
\begin{itemize}
\item $\Delta t\leq 1$ day, in order to sample a large number of returns in each time series and to exclude intra-night returns.
\item $\delta t \gg 1$ second (i.e. the time resolution of our datasets), in order to avoid micro-structural noise and to measure significant fluctuations of stock prices. We fix $\delta t \geq 1$ minute in D1 and $\delta t \geq 10$ seconds in D2.
\item The ratio $\Delta t/\delta t$, which is equal to the number of variables $N$, should be large enough to have a statistically significant measure of the IPR, but should not be too large, because the frequency of large deviations would become too small. We fix $N$ in the range $20\leq N \leq100$.
\end{itemize}

It is well-known that intra-day returns exhibit a larger volatility at the opening and at the closure of markets than in the rest of the day. In order to avoid these effects of market seasonality, we remove from the time series the first and the last 30 minutes of trades of each day. Furthermore, datasets show anomalous periods where trades are temporary absent. The most common examples are the ``half working days'' in the American market and the volatility auctions in the Italian market. We exclude from our analysis all returns occurring during those periods.

Large deviations of price returns are rare events, so we need very long time series in order to observe the main features of their realizations. Generally, the results obtained from a single stock are not enough to statistically analyze the large deviations of price returns (see \ref{app:findata_APP}). In order to overcome this problem, we merge the results obtained from different stocks in a single output (one for each dataset). Since different stocks usually exhibit different price fluctuations, this merging requires a proper rescaling of the measured observables.
In our analysis, we rescale the stock returns $R$ according to the \emph{stock's volatility} $\sigma$, which is the typical price variation for a specific stock. We evaluate $\sigma$ as the root mean square of the total returns $R$ over the whole time-series (the obtained value for $\sigma$ depends on the chosen time-scale $\Delta t$).

\subsection{Results}

\begin{figure}
\centering
\includegraphics[width=\textwidth]{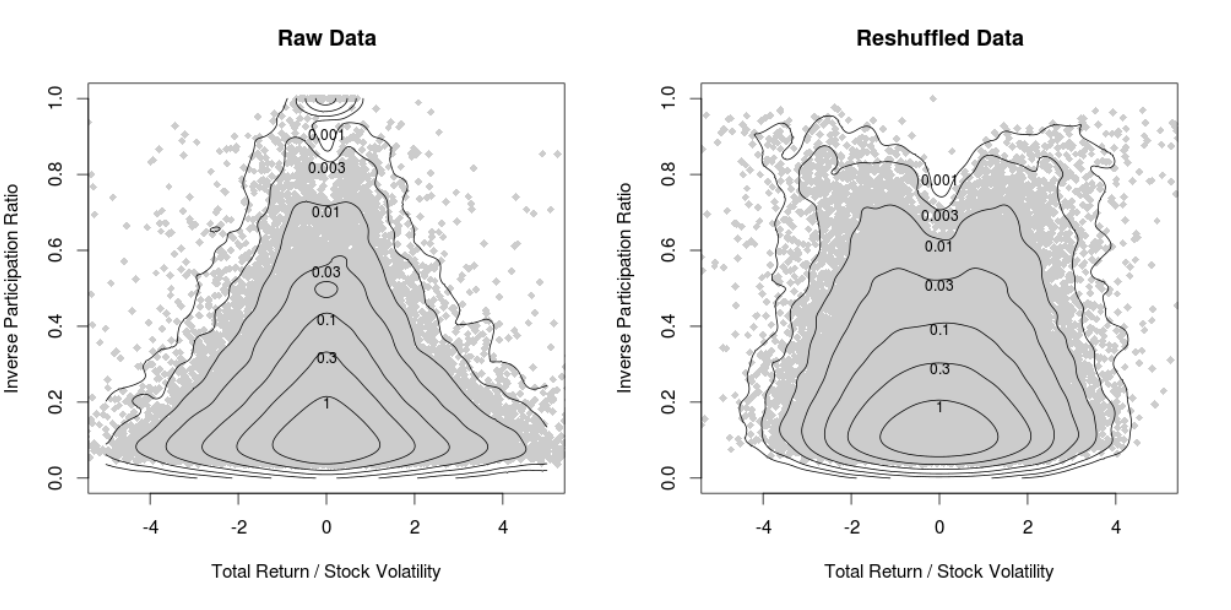}
\caption{Scatter-plots of $I_2$ vs $R/\sigma$ for all 40 stocks in D2, for $\Delta t=30$ minutes and $\delta t=30$ seconds ($N=60$), obtained from the raw (left) and reshuffled (right) time series. The iso-probability contours of the joint probability density of $R$ and $I_2$ are shown (the density is evaluated through a Gaussian kernel estimation with bandwiths $(0.2,0.02)$).}
\label{FIG:MAINPLOT}
\end{figure}

In Fig.\ \ref{FIG:MAINPLOT} we show the results obtained in a specific case: it is the scatter plot of $I_2$ versus $R/\sigma$, obtained from the analysis of all 40 stocks in D2 with $\Delta t = 30$ minutes and $\delta t = 30$ seconds ($N=60$).
Since the assumption {\bf B} of independent returns is not empirically verified,
we repeat the same observations on a reshuffled dataset and we compare the results. The reshuffled dataset has been obtained by rearranging the time series of price returns of each stocks in a random order, in this way we preserve the statistical properties of single returns but we destroy their dependency over time.

Fig.\ \ref{FIG:MAINPLOT} shows a significant difference between raw and reshuffled data. In reshuffled data, large returns can have very high IPR, therefore large deviations tend to concentrate, and our expectation is confirmed (compare Fig.\ \ref{FIG:MAINPLOT} with the Student's case in Fig.\ \ref{FIG:SIMULATION}). On the other hand, in raw data, large returns exhibit much lower values of the IPR, and the largest returns are almost always related to a very low IPR (close to its minimum value). As a result, concentration phenomena are inhibited by time-dependence effects, and large deviations of price returns tend to be equally distributed in time.

Since both raw and reshuffled time series are obtained from the same sequence of small-scale returns $r_i$ (except for their ordering), the inhibition of concentration phenomena implies that stock prices somehow react to (or prepare for) extreme price jumps. More specifically, extreme occurrence of $r_i$ in actual time series are usually followed or preceded by untypical price fluctuations that reduce the final values of $R$ (jump reversion) or reduce the final value of $I_2$ (volatility increase).

The presented result is non-trivial and very counterintuitive. Generally, one is tempted to explain large deviations of stock prices as the effect of extreme price jumps. Fig.\ \ref{FIG:MAINPLOT}, instead, clearly shows that discontinuous price jumps (high-IPR points) generate smaller returns than continuous price fluctuations (low-IPR points). Our observations suggest that the propagation of extreme financial events from small to large time-scales is affected by two opposite dynamics: a \emph{reduction feedback}, which reduces the effects of jumps and irregularities in price movements, and an \emph{amplification feedback}, which amplifies the regular diffusion of prices.

All these considerations have been drawn from the observation of Fig.\ \ref{FIG:MAINPLOT}, corresponding to a specific dataset, but our results seem to be much more general. Indeed, by investigating both D1 and D2 for different choices of the time steps $\Delta t$ and $\delta t$, we have been able to observe the same phenomenon on different markets, periods, and time-scales. Specifically, the inhibition of concentration phenomena in large returns due to time-dependence effects is not peculiar of high-frequency returns, but can be observed also on daily time-scales. The same behavior is observed also for different choices of the parameter $k$ of the IPR $I_k$. These results are reported in \ref{app:findata_APP}.

\subsection{On the role of volatility}

The reduction/amplification feedbacks that we recognized in the previous section could be explained in terms of {\em volatility clustering}. As we already pointed out, the typical scale of price returns is highly correlated in time, even though returns themselves are not. Therefore, price returns with similar scale tend to cluster in time, giving rise to periods of variable length where the volatility is approximately constant. Interestingly, such periods are not characterized by a typical time-scale: the duration of the volatility clusters spans several orders of magnitude, and this allows to observe them at any frequency. High volatility periods, for instance, can last few seconds as well as several days, during crisis periods. Such phenomenon, also known as {\em volatility intermittency}, is usually invoked as an evidence of the {\em multi-fractal nature} of stock prices' returns \cite{MRW_bouchaud}.

We can suppose that, during constant volatility periods, price returns are much less fat-tailed than expected from a global analysis, and that their overall fat-tailed nature arises from mixing different periods with heterogeneous volatilities. Within this picture, we can try to explain the reduction/amplification feedbacks as a consequence of the volatility intermittency. If large returns occur only during high volatility periods, than they should be accompanied by other returns of the same scale (with random signs). Such returns may interfere in a disruptive or constructive way, giving rise to the reduction and amplification feedbacks we noted above. A constructive interference may cause a large final return (large $|R|$), but is perceived as a diffusive fluctuation rather than a sharp jump (low IPR). Moreover, the absence of a typical time-scale for the duration of the volatility clusters explains why we are able to observe the inhibition of concentration phenomena on several frequencies.

In order to test this explanation, we compare our measurements on financial time series with new ones on a simulated dataset, which has been obtained through the so-called {\em multi-fractal random walk} \cite{MRW_bouchaud,MRW_bacry}. This stochastic process has been introduced by Bacry, Delour and Muzy in 2001 \cite{MRW_bacry} to mimic the volatility intermittency of stock prices' returns and its multi-fractal nature. It is defined by two main parameters: the {\em integral time-scale $T_\mathrm{int}$} and the {\em intermittency coefficient $\lambda^2$}. At a time-scale larger than $T_\mathrm{int}$ the increments of the multi-fractal random walk are independent, but below $T_\mathrm{int}$ they exhibit a specific multi-fractal behavior which is regulated by $\lambda^2$. The larger $\lambda^2$, the stronger is the intermittent behavior of the volatility. The marginal distribution of returns turns out to be a power-law distribution for all strictly positive values of $\lambda^2$ (with a tail exponent equal to $\lambda^{-2}$). For $\lambda^2\to0$ the multi-fractal random walk looses its multi-fractal nature and reduces to a simple Browian motion.

\begin{figure}
\centering
\includegraphics[width=\textwidth]{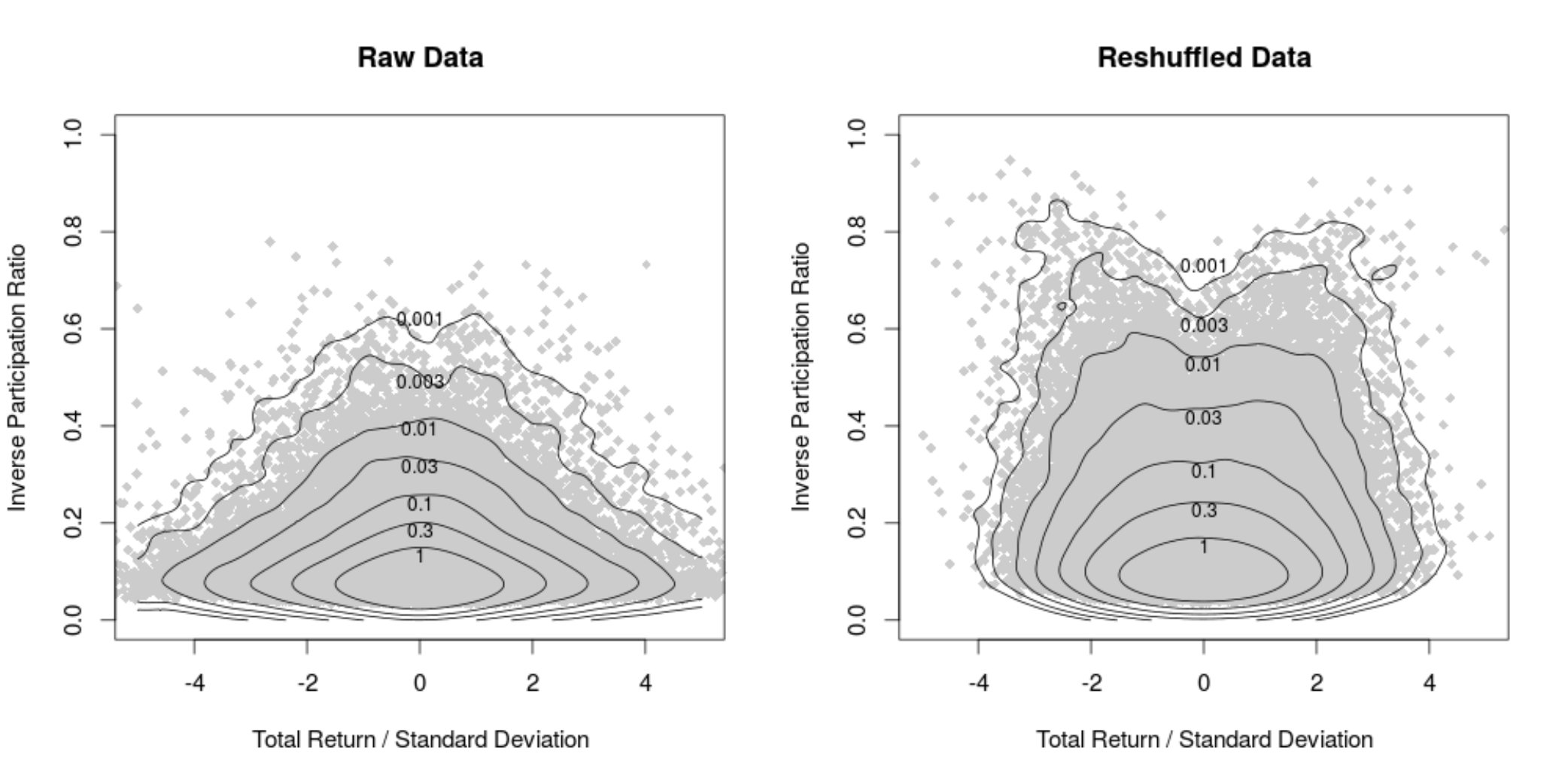}
\caption{Scatter-plots of $I_2$ vs $R$ for a multi-fractal random walk with $\lambda^2=0.03$ and $T_\mathrm{int}=10^5$, obtained from a raw (left) and a reshuffled (right) time series. The plots are obtained from 213,000 observations of $N=60$ returns, corresponding to the same number of observations performed in Fig.\ \ref{FIG:MAINPLOT}. The iso-probability contours of the joint probability density of $R$ and $I_2$ are shown (the density is evaluated through a Gaussian kernel estimation with bandwiths $(0.2,0.02)$).}
\label{FIG:MULTIFRACTAL}
\end{figure}

In Fig.\ \ref{FIG:MULTIFRACTAL} we redraw the scatter-plots $I_2$ vs.\ $R$ of Fig.\ \ref{FIG:MAINPLOT} using the simulated data obtained from a multi-fractal random walk with $\lambda^2=0.03$ and $T_\mathrm{int}=10^5$ (in units of $\delta t$).
The new scatter-plots reproduce quite well the ones obtained from financial data, showing the same inhibition of concentration phenomena and the emergence of the reduction/amplification feedbacks. The only visible difference with the previous plots is due to a general decreasing of the IPR in the time-ordered series (left plot). The parameter $\lambda^2$ has been chosen accordingly to the empirical estimate reported in \cite{MRW_bouchaud}, based on the multi-fractal scaling laws of the moments of returns. After verifying that the p.d.f.\ of the simulated returns recovers the shape of the experimental one, we notice that by comparing the plots in Figs.\ \ref{FIG:MAINPLOT} and \ref{FIG:MULTIFRACTAL} the qualitative behaviour of the IPR is also reproduced.
As long as $T_\mathrm{int}\gg N$, the results are not sensitive to the specific value of $T_\mathrm{int}$, which is the only time-dimensional parameter of the multi-fractal random walk. Our analysis shows that $T_\mathrm{int}$ can be changed by orders of magnitude without altering too much the shape of the scatter-plot. This is in agreement with the observation on real data, where the inhibition of concentration phenomena can be observed across different time-scales.

In conclusion, our measurements on real financial data, showing the inhibition of concentration phenomena in the returns of stock prices and a non-trivial propagation of rare events from short to large time-scales, can be justified thanks to the intermittent behavior of the volatility and to the multi-fractal nature of the price-fluctuation process.

\section{Financial correlations and the market mode as a large deviation}
\label{sec:marketmode}

In Sec.\ \ref{sec:ldrmt} we discussed how large deviations in Wishart-like random matrices with fat-tailed independent elements may concentrate on the largest eigenvalue. Now it is tempting to apply such results to covariance matrices of financial stock returns. Indeed, financial covariance matrices exhibit a clear concentration phenomenon, where the largest eigenvalue -- the so-called {\em market mode} -- is widely separated from the rest of the distribution and it roughly scales with the number $N$ of stocks \cite{Laloux,Plerou}.

The emergence of the market mode is usually explained as the consequence of a common factor, the market itself, which linearly couples to all other stocks. Such {\em one-factor model} is generally enough to explain both the scaling-law of the concentrated eigenvalue and the shape of the corresponding eigenvector. This description, though appealing, assumes a specific mechanism for the dynamics of the system, which has its shortcomings. For example, the factor depends on the universe of assets considered. Here we show that the market mode can be explained as a large deviations phenomenon, without assuming any specific mechanism. While the factor model is one way in which the large deviation can be realized, the large deviation approach is not based on any assumptions on specific mechanisms or models, thus allowing a more general explanation for the emergence of the market mode. Moreover, any observation of a system of interacting variables can be treated in principle as a large deviating sample of some i.i.d.\ variables; this approach allows to reinterpret the system’s interactions as the deviation of some observable, setting the problem in a different light.

As we discussed in Sec.\ \ref{sec:ldrmt}, the concentration of the largest eigenvalue occurs when:
\begin{description}
  \item[A] returns have a fat tailed distribution $Q$ (see Fig.\ \ref{FIG:PDF});
  \item[B] the mean of the eigenvalues attains an atypically large value: $\frac{1}{n}\sum_i \lambda_i=\bar\lambda>\langle \lambda\rangle_Q$.
\end{description}
In financial jargon, statement {\bf B} is akin to {\em excess volatility}, since $\sum_i \lambda_i={\rm Tr}~\hat{M}$ is the sum of the variances of stock returns, which are usually called volatilities.

As we have seen in Sec.\ \ref{sec:ldrmt}, the consequences of {\bf A} and {\bf B} are that:
\begin{description}
  \item[C] the largest eigenvalue is widely separated from the rest of the distribution and it is proportional to $N$;
  \item[D] the eigenvector $\ket{\lambda_1}$ corresponding to the largest eigenvalue is localized.
\end{description}

While empirical covariance matrices of financial returns exhibit property {\bf C}, they fail to satisfy {\bf D}: the eigenvector $\ket{\lambda_1}$ is an extended one, i.e.\ it has all positive and roughly constant components \cite{Plerou2,RaffaelliMarsiliPonsot}. Therefore, the hypotheses of fat tailed returns {\bf A} and of excess volatility {\bf B} -- i.e. of a large deviation in return volatility ${\rm Tr} ~\hat M$ -- are not the correct hypotheses to explain the emergence of a large eigenvalue in the spectrum of the covariance matrix $\hat M$. 

Furthermore, the analysis of Section \ref{sec:ldrmt} predicts that the large deviation can be traced back to an anomalously large return in one stock $i^*$ and at one time $t^*$, with $|x_{i^*,t^*}|\sim \sqrt{NT}$, whereas the largest  $N\times T$ i.i.d.\ draws from a distribution with power law fat tail $q(x)\sim |x|^{-\gamma-1}$ is of the order $\max_{i,t} |x_{i,t}|\sim (NT)^{1/\gamma}$, which is much smaller ($\gamma\approx 3$ in financial data). Hence even {\bf A} is violated by at least one point $x_{i^*,t^*}$ in the sample, which is non-typical.
As a matter of fact, the large deviation scenario above would predict localization in $t$ also, i.e.\ that the largest eigenvector of the matrix $D_{t,t'}=\frac{1}{N}\sum_ix_{i,t}x_{i,t'}$ is also localized. For financial data we don't find such a sharp localization.

What is the simplest set of observations which explains the emergence of a market mode in financial correlations?
We argue below that the answer to this question is {\bf A} and 
\begin{description}
  \item[E] {\em excess covariance}: 
\begin{equation}
\label{LDtotcorr}
\bra{1}{\hat{M}}\ket{1}=\sum_{i,j}M_{i,j}=N^2c > \left\langle \bra{1}{\hat{M}}\ket{1}\right\rangle_Q \ ,
\end{equation}
where $\ket{1}=(1,1,\ldots,1)$.
\end{description}
The typical value of the total covariance is of order $N$. Indeed note that
\begin{equation}
\label{Cexpansion}
\bra{1}{\hat{M}}\ket{1}=\sum_{i=1}^N\lambda_i|\braket{1}{\lambda_i}|^2 \ .
\end{equation}
In the Wishart ensemble $\lambda_i\sim O(1)$ and, because of rotational invariance, the eigenvectors $\ket{\lambda_i}$ can be considered as random vectors on the unit sphere. Since $\braket{\lambda_i}{\lambda_i}=1$, each component of $\ket{\lambda_i}$ is of order $1/\sqrt{N}$ (with a random sign) and the projection $\braket{1}{\lambda_i}$ of the eigenvectors on the vector $\bra{1}$ is also of order one. Therefore Eq.\ (\ref{Cexpansion}) is a sum of $N$ positive terms of order one, i.e.\ it is of order $N$. For non-Wishart ensembles the rotational invariance is broken, yet, as long as $\gamma>2$, it is reasonable to assume that $\bra{1}{\hat{M}}\ket{1}/N$ attains a typical value which is the same as the one in the Wishart ensemble.

It is important to notice that the existence of a linear market factor naturally implies {\bf E}. Indeed, when the common factor $f_t$ is included to the system through the substitution $x_{i,t} \mapsto x_{i,t} + b_i f_t$, the total covariance becomes
\[ \bra{1}{\hat{M}}\ket{1} \mapsto \bra{1}{\hat{M}}\ket{1} +\cdots+ \sigma_f^2\sum_{i,j}b_ib_j\ , \]
where $\sigma_f^2=\sum_{t=1}^T f_t^2/T$. If $b_i\approx\mathrm{constant}$, as we expect for a market-like factor, then the last term of this expression becomes proportional to $N^2$. On the contrtary, since {\bf E} does not imply the existence of a market factor, the hypothesis {\bf A} and {\bf E} could provide a more general explanation for the emergence of the market mode than the usual one-factor model.

A large deviation where the total covariance is of order $N^2$, as in Eq.\ (\ref{LDtotcorr}), can be achieved by {\em i)} aligning the eigenvector corresponding to the largest eigenvalue to $\ket{1}$, so that $|\braket{1}{\lambda_1}|^2\sim O(N)$ and by {\em ii)} having the largest eigenvalue $\lambda_1\sim O(N)$ (i.e. {\bf C}). Loosely speaking, this is what we expect to occur since, {\em i')} because of rotation invariance, aligning $\ket{\lambda_1}$ to $\ket{1}$ has no ``cost'' in terms of free energy (i.e.\ in terms of the logarithm of the probability of the deviation) in the Wishart ensemble. Fat tails are expected to break rotation invariance, so there might be a cost associated with the rotation. Furthermore {\em ii')} concentrating a large deviation on the largest eigenvalue has the minimal free energy cost for fat tailed distributions. 
Our aim is to make this intuition precise and to show that both {\em i)} and {\em ii)} typically occur when we enforce {\bf A} and {\bf E} with $c$ taking the same value as for financial data.

We derived a procedure to sample ensembles with fixed total covariance, drawing $x_{i,t}$ from a given underlying distribution $Q$. While interesting in its own, a detailed discussion of this method and the results will lead us too far. In brief, it turns our that imposing a constraint on the total covariance as in {\bf E} distorts the underlying distribution $Q$ of the random variables, i.e. it invalidates {\bf A}. We refer the interested reader to \ref{app:excesscov_APP} for a detailed discussion. 

Here we concentrate on a sampling scheme that allows us to preserve {\bf A} while enforcing {\bf E}. The idea is to perform a Montecarlo based on reshuffling of the elements $x_{i,t}$, by changing the time order of different returns. These moves are accepted with probabilities that depend on how the total covariance $\bra{1}{\hat{M}}\ket{1}$ changes. These moves clearly leave the distribution of returns unchanged. More precisely, we perform a \emph{reshuffling}-Montecarlo simulation based on the following steps:
\begin{enumerate}
\item Generate a random time series $x_{i,t}$ according to the p.d.f.\ $p(x)$. Then define the local times $\tau_k(t)$ and set them to $\tau_k(t)=t$ for all $k$.
\item Pick an index $k$ and two times $t\neq t'$ at random. 
\item Swap the values of the local times corresponding to $t$ and $t'$ for index $k$, i.e.\ set $(\tau_k(t),\tau_k(t'))\to (\tau_k(t'),\tau_k(t))$.
\item Compute
\[ M_{i,j}=\frac{1}{T}\sum_{t=1}^T x_{i,\tau_i(t)}\,x_{j,\tau_j(t)} \ . \]
\item Accept the move with probability $p=e^{-\beta \Delta \mathcal{H}}$ where $\Delta \mathcal{H}$ is the change in the Hamiltonian
\[ \mathcal{H}=-\bra{1}\hat M\ket{1} \ , \]
otherwise reject it.
\item Repeat steps (ii)--(v) until a stationary state is reached.
\end{enumerate}
The above algorithm has the interesting advantage that it does not require the knowledge of $p(x)$ and it can be run directly on empirical data. Instead of generating a random time series in step (i), $x_{i,t}$ can be taken directly from financial data. The algorithm preserves the marginal distribution, thus it fixes the total volatility ${\rm Tr}~\hat M$ to its starting value. Upon changing the value of $\beta$, we expect that this algorithm samples different ensembles of time series with a given total correlation $\bra{1}\hat M\ket{1}\simeq N^2 c(\beta)$, where $c(\beta)$ is an increasing function of $\beta$. The typical case is obtained with $\beta=0$.

\begin{figure}
\includegraphics[width=\textwidth]{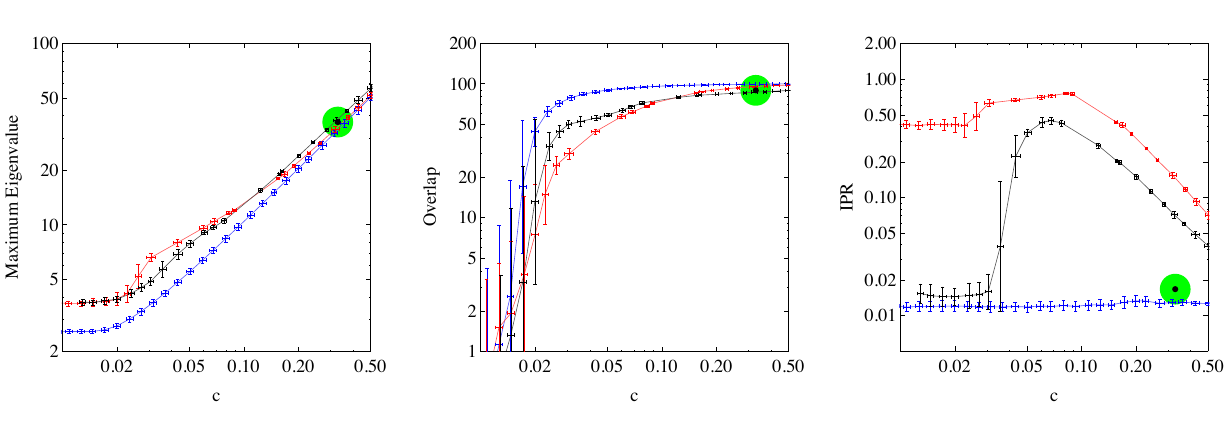}
\caption{
Ensembles with excess covariance at fixed marginal distribution.
From left to right, as a function of $c$, we plot 
$\lambda_1$ (left), the overlap $|\braket{1}{\lambda_1}|^2$ (center) and the Inverse Participation Ratio (IPR) of 
$\braket{x_t}{\lambda_1}$ (right), which measures the degree of concentration of the mode $\ket{\lambda_1}$ over time. The latter is defined as $\mathrm{IPR}=\sum_t \braket{x_t}{\lambda_1}^4/(\sum_t\braket{x_t}{\lambda_1}^2)^2$, where $\ket{x_t}$ is the vector whose $i^{\rm th}$ component is $x_{i,t}$. We expect $\mathrm{IPR}\sim 1/T$ when $\braket{x_t}{\lambda_1}\simeq\mathrm{const}$ is completely delocalized in $t$ and 
$\mathrm{IPR}\sim 1$ when $\braket{x_t}{\lambda_1}$ is a completely concentrated on just one value of $t$.
In all plots, we report results for the Normal distribution (Blue), Student's t-distribution with 4 degrees of freedom (Red), and the empirical distribution of price returns from D1 (Black).
The dot in the green circle denotes the measurements on the original financial data.
The results correspond to the financial time series of $N=100$ stocks over $T=249$ days.
}
\label{FIG:MC_B}
\end{figure}

We executed the above procedure both on real and simulated time series. The simulated series have been drawn from a normal and a Student-t distribution, whereas the empirical series were obtained from the dataset D1 (see the previous section) and corresponds to the daily return of the 100 most traded stocks of the New York Stock Exchange in the year 2003 ($N=100$, $T=249$; daily returns have been evaluated as the difference of log-prices at market's closure). The empirical distribution of returns is shown in Fig.\ \ref{FIG:PDF} and has been rescaled in order to have the same variance of the simulated returns.

As $\beta$ increases the ensemble draws a curve in the $(\lambda_1(\beta),c(\beta))$ plane. In each observed case, this curve passes very close to the point $(\lambda_1^\mathrm{market},c^{\rm market})$ corresponding to the original financial data
(see Fig.\ \ref{FIG:MC_B}). Contrary to what happens for large deviations at fixed ${\rm Tr}~\hat M$, where the large deviation is localized both across stocks $i$ and in time $t$, Fig.\ \ref{FIG:MC_B} shows that large deviations with excess covariance $\bra{1}\hat M\ket{1}$ are neither localized on a single stock nor in a single day.  
The value of $|\braket{1}{\lambda_1}|^2$ comes very close to those observed in real market data, whereas the IPR measuring concentration in $t$ is much smaller than one, though it's not as small as in real market data. 
This allows us to conclude that the market mode can be actually explained as a large deviation of the excess covariance. 

\begin{figure}
\begin{center}
\includegraphics[width=0.5\textwidth]{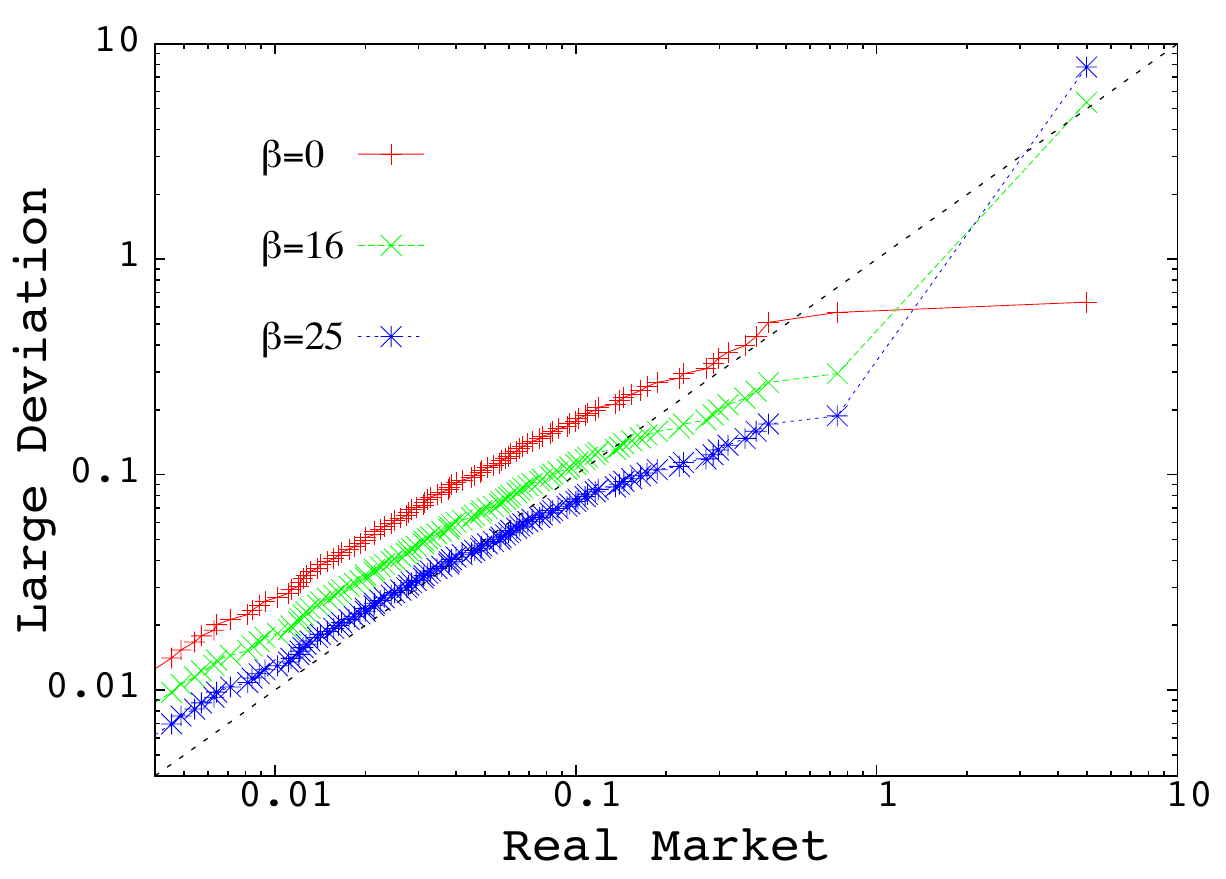}
\end{center}
\caption{The value of the $i^{\rm th}$ largest eigenvalue in the large deviation $\lambda_i^{(LD)}$ versus the corresponding eigenvalue $\lambda_i^{({\rm market})}$ of the true market covariance matrix for $\beta=0$ that reproduces the Wishart spectrum, for $\beta=16$ that corresponds to the best fit to market data, and for $\beta=25$.}
\label{FIG:MC_C}
\end{figure}

Fig.\ \ref{FIG:MC_C} shows that this does not explain the whole structure of the covariance matrix. The figure reports the value of the $i^{\rm th}$ largest eigenvalue $\lambda_i^{\rm (LD)}$ in the large-deviation ensemble versus the corresponding eigenvalue $\lambda_i^{({\rm market})}$ of the true market covariance matrix.
As $\beta$ increases, the largest eigenvalue $\lambda_1^{\rm (LD)}$ takes more and more of the total sum ${\rm Tr} \hat M$ (we remind that the ${\rm Tr} \hat M$ is constant in the Montecarlo procedure). Yet, even when $\lambda_1^{\rm (LD)}$ is adjusted such that $\lambda_1^{\rm (LD)} \approx \lambda_1^{\rm (market)}$, the other $N-1$ eigenvalues in the large-deviation ensemble significantly deviate from the real ones. Specifically, in all plotted cases, the distribution of real eigenvalues turns out to be broader than the one obtained from the large-deviation ensemble.

\section{Conclusions}

The phenomenon discussed in this paper, that large deviations for fat tailed distributions concentrate, has been discussed e.g. in \cite{Embrechts} for risk models and \cite{satya,satya_2} for mass transport models. 
We believe this phenomenon is of general relevance and has not received enough attention. First, because fat tailed distributions are widespread in a large variety of phenomena. Second, because this phenomenon is one of a textbook realizations of a second order phase transition. Third, because it has unintuitive consequences for inference, for large deviations of stochastic processes and random matrix theory, as we have argued. 

Concentration is typically perceived as a non-trivial phenomenon which begs for some explanation. For example, economic geography \cite{Krugman} has gone a long way trying to explain why economic activity concentrates on the same geographic location. Yet if one acknowledges that economic activity (e.g. firm size) is generally fat tailed distributed, what really needs to be explained is not why economic activity concentrates but rather why it does not concentrate as much as we should expect. As we have seen, the same applies to returns of financial assets: there what needs to be explained is why there are not as many jumps as one would expect when large excursions occur.

It is tempting to speculate on the possible application of these results to biological evolution. We think of evolutionary processes as occurring by the accumulation of mutations on the genome. Surviving individuals are those that achieve fitness changes that are large enough. The effects of a mutation on the fitness is very complex and in general non-linear, but neglecting epistatic effects, one can consider the fitness change as the sum of the effects of individual mutations, in a zeroth order approximation. Then if fitness changes of individual mutations have a broad distribution, one is lead to the conclusion that fit species are not likely to result from the accumulation of small positive mutations. Rather they are likely to arise from large fitness jumps, which is somewhat reminiscent of the notion of punctuated equilibria \cite{Gould} as contrasted to phyletic gradualism.

Our discussion has no further pretense than to illustrate how the concentration of large deviations for fat tailed distribution may lead to counterintuitive results, showing that phenomena such as sharp changes or strongly uneven fluctuations can arise as a result of pure randomness, 
without having to invoke any specific mechanism. In the terminology of Ref. \cite{DragonKings}, Dragon Kings typically occur in large deviations with fat tailed distributions. Indeed, as in other more complex phenomena (e.g. phase transitions), pure randomness here manifests through a {\em symmetry breaking} phenomenon, whereby the {\em a priori} equivalence of the data points in the sample is spontaneously broken. Given the widespread occurrence of fat tailed distributions, this is likely to be an important fact of chance to take into account. 

\section*{Acknowledgements}

Financial support from F.S.E.\ within the framework of the S.H.A.R.M.\ P.O.R.\ 2007/2013 project, from LIST S.p.A.\, and from the NETADIS Marie Curie Training Network of the European Commission (FP7 -- Grant 290038) is acknowledged.
We acknowledge the M.I.U.R.\ research project {\em ``Dinamica di altissima frequenza nei mercati finanziari''} for providing dataset D1 and LIST S.p.A. for providing dataset D2.
The authors want to thank J.-P. Bouchaud, L.\ Caniparoli, S. N. Majumdar, and D. Sornette for fruitful discussions, E.\ Dameri and E.\ Melchioni for fostering the present collaboration, and D.\ Davio for the continuous encouragement.

\appendix

\section{Perturbation theory for large deviations of random matrices with fat tailed distributions}
\label{app:perturbation_theory}

We take as unperturbed system $H_0$ and as perturbation $V$ the matrices
\[ H_0=
\left(\begin{array}{cc}\lambda_0 & 0 \\ 0 & \hat C\end{array}\right),\qquad
V=\left(\begin{array}{cc}0 & \bra{b} \\\ket{b} & 0\end{array}\right) \ , \]
where $\lambda_0$, $\ket{b}$, and $\hat C$ are, in order, a scalar, a $(N-1)$ vector, and a $(N-1)\times (N-1)$ matrix with components
\[ \lambda_0=\frac{1}{T}\sum_{t=1}^T x_{0,t}^2 \ , \qquad b_i=\frac{1}{T}\sum_{t=1}^T x_{0,t}x_{i,t} \ , \qquad C_{ij}=\frac{1}{T}\sum_{t=1}^T x_{i,t}x_{j,t} \ . \]
We denote the eigenvalues and the eigenvectors of the matrix $\hat C$ as $\lambda_n$ and $\ket{\lambda_n}$, respectively.
Hence, the complete set of eigenvectors of $H_0$ is given by
\[ \ket{\Lambda_n}=\left\{\begin{array}{cc}\ket{1,0} & \hbox{with eigenvalue~} \lambda_0 \ , \\\ket{0,\lambda_n} & \hbox{with eigenvalue~} \lambda_n \ . \end{array}\right. \]
As in Sec.\ \ref{sec:EIGEN_RM}, we focus on the case where $\lambda_0\simeq O(N)$ and the sums over the elements $x_{i,t}$ are dominated by the single term $x_{0,t^*}\simeq \sqrt{T\lambda_0}$, whereas all others $x_{i,t}$ for $i\neq0$ or $t\neq t^*$ are i.i.d.\ random variables with zero mean and finite variance $\sigma^2$.
In this case we can write
\[ b_i=\frac{x_{0,t^*}}{T}x_{i,t^*}+\frac{1}{T}\sum_{t\neq t^*} x_{0,t}x_{i,t} \simeq\sqrt{\frac{\lambda_0}{T}}x_{i,t^*}+\frac{\sigma^2}{\sqrt{T}}\zeta_i \ , \]
where, by  the Central Limit Theorem, $\zeta_i\sim\mathcal{N}(0,1)$ is a Gaussian variable with zero mean and unit variance. Let $v_{n,i}$ be the $i^{\rm th}$ component of $\ket{\lambda_n}$ and assume that $v_{n,i} \sim\mathcal{N}(0,1/N)$. Then
\[ \braket{b}{\lambda_n}=\sum_{i=1}^{N-1}b_i v_{n,i}\simeq \sqrt{\frac{\lambda_0}{T}}\sigma\left(1+\frac{\sigma}{\sqrt{\lambda_0}}\right)\eta_n \ , \]
where $\eta_n\sim\mathcal{N}(0,1)$.
Up to the first two orders in perturbation theory, the correction to the eigenvalues is given by
\begin{eqnarray*}
\lambda_0' & = & \lambda_0+\sum_{n=1}^{N-1} \frac{|\braket{b}{\lambda_n}|^2}{\lambda_0-\lambda_n} \ , \\
\lambda_n'&=& \lambda_n-\frac{|\braket{b}{\lambda_n}|^2}{\lambda_0-\lambda_n} \ ,
\end{eqnarray*}
where the first order correction vanishes, since $V$ has zero diagonal matrix elements. The eigenvectors, up to the first order, read
\begin{eqnarray*}
\ket{\Lambda_0'} & = & \ket{1,0}+ \sum_{n=1}^{N-1} \frac{\braket{b}{\lambda_n}}{\lambda_0-\lambda_n}\ket{0,\lambda_n} \ , \\
\ket{\Lambda_n'} & = & \ket{0,\lambda_n}- \frac{\braket{b}{\lambda_n}}{\lambda_0-\lambda_n}\ket{1,0} \ .
\end{eqnarray*}
For $N,T\to\infty$, at fixed $q=T/N$, the distribution of the eigenvalues $\lambda_n$ tends to the Mar\v cenko-Pastur distribution $\rho(\lambda)d\lambda$ \cite{BurdaJurkiewicz}. Therefore, for large values of $\lambda_0\sim N\sim T$, the correction to $\lambda_0$ can be estimated as
\[ \lambda_0' \simeq \lambda_0+\frac{ \lambda_0\sigma^2}{T/N}\int\!d\lambda\frac{\rho(\lambda)}{\lambda_0-\lambda}\simeq   \lambda_0+\frac{\sigma^2}{q}+O(1/\lambda_0) \ , \]
whereas $\ket{\Lambda_0'}$ remains close to $\ket{1,0}$ and is still a localized vector.

\section{Additional results about concentration of large returns in financial time series}
\label{app:findata_APP}

In this section we present some additional results about concentration phenomena in financial time series, following the analysis of Sec.\ \ref{sec:findata}. All results are presented as scatter plots of $I_k$ vs.\ $R$, as performed in Fig.\ \ref{FIG:MAINPLOT}.

\begin{figure}
\centering
\includegraphics[width=\textwidth]{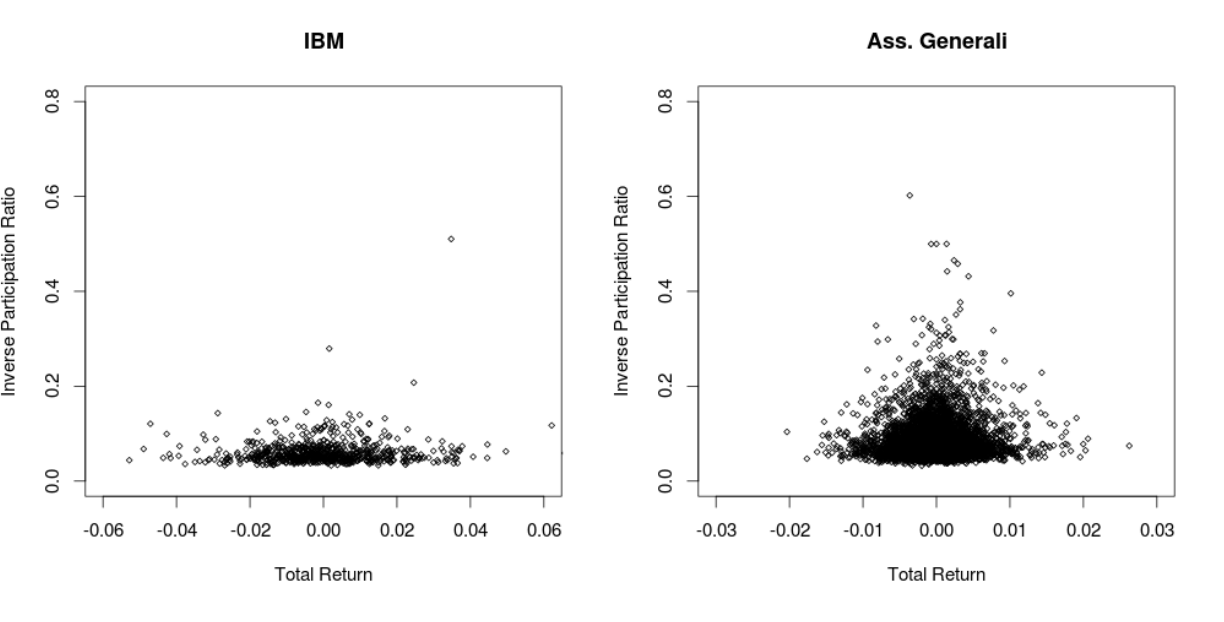}
\caption{Scatter-plots of $I_2$ vs $R$ for two sample stocks: IBM (from D1)
for $\Delta t=1$ trading day (5.5 hours) and $\delta t=5$ minutes ($N=66$) (left); and Assicurazioni Generali (from D2) for $\Delta t=30$ minutes and $\delta t=30$ seconds ($N=60$) (right).}
\label{FIG_SUPP:EXAMPLES}
\end{figure}

In Fig.\ \ref{FIG_SUPP:EXAMPLES} we show the scatter plots for two sample stocks taken from D1 and D2, respectively. Plots show the typical result obtained from a single stock and highlight how the shape of the figures can change from stock to stock. The points corresponding to rare events are too sparse and do not allow a reliable analysis of concentration phenomena. Hence, instead of single stocks, we always analyze the superposition of all stocks in a given dataset, after a proper rescaling of stock's returns in terms of stock's volatility.

\begin{figure}
\centering
\includegraphics[width=0.8\textwidth]{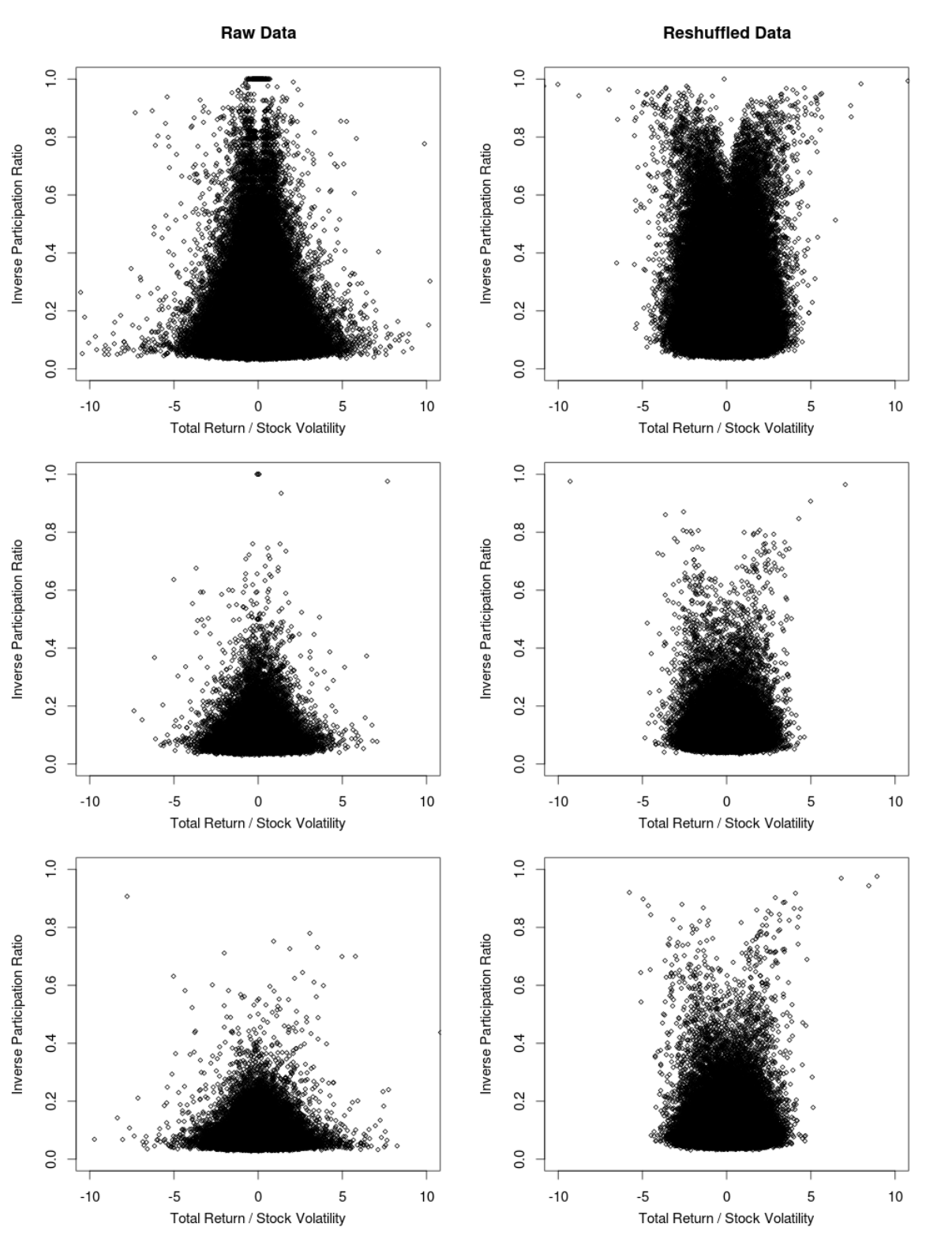}
\caption{Scatter-plots of $I_2$ vs $R/\sigma$ for different datasets and time-scales. Top: all 40 stocks in D2, for $\Delta t=30$ minutes and $\delta t=30$ seconds ($N=60$). Center: all 40 stocks in D2, for $\Delta t=150$ minutes and $\delta t=150$ seconds ($N=60$). Bottom: all 100 stocks in D1, for $\Delta t=1$ trading day (5.5 hours) and $\delta t=5$ minutes ($N=66$). Plots are shown for both raw (left) and reshuffled (right) time series.}
\label{FIG_SUPP:D1D2}
\end{figure}

In Fig.\ \ref{FIG_SUPP:D1D2} we report again the main result shown in Fig.\ \ref{FIG:MAINPLOT} (top), together with similar results obtained from different datasets and time-scales (center and bottom). Due to the larger time-scale, the new scatter-plots are composed of a reduced number of points. Yet, they exhibits the same features observed in Fig.\ \ref{FIG:MAINPLOT}, such as the inhibition of concentration phenomena for large returns and the amplification (reduction) of continuous (discountinuous) price movements. It is worth to recall that the new scatter-plots refer not only to different time-scales, but also to different markets and periods. This proves how the above phenomena are very general and do not depend on the specific dataset used for our observations.

\begin{figure}
\centering
\includegraphics[width=0.8\textwidth]{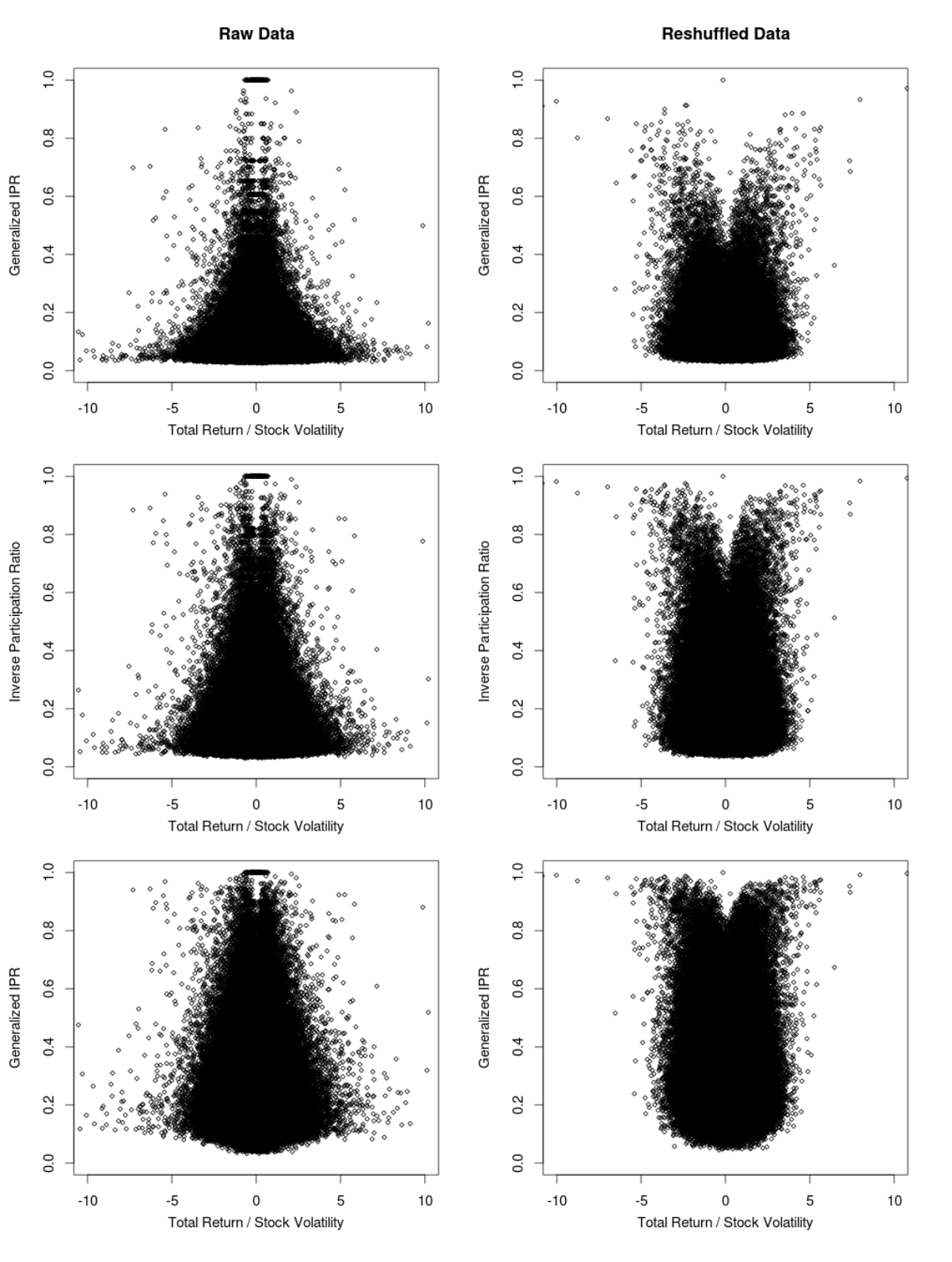}
\caption{Scatter-plots of $I_k$ vs $R/\sigma$ for the same data of Fig.\ \ref{FIG:MAINPLOT} for different values of $k$. Top: $k=1$. Center: $k=2$. Bottom: $k=\infty$. Plots are shown for both raw (left) and reshuffled (right) time series.}
\label{FIG_SUPP:IPRS}
\end{figure}

Finally, in Fig.\ \ref{FIG_SUPP:IPRS}, we re-plot Fig.\ \ref{FIG:MAINPLOT} for different parameters $k$ of the IPR $I_k$. Here we show the two limits $k\to1$ and $k\to\infty$, namely:
\[ I_1 =\lim_{k\to1} I_k = \exp \left ( \sum_{i=1}^N w_i \ln w_i \right ) \ , \]
and
\[ I_\infty =\lim_{k\to\infty} I_k = \max_{i=1\dots N}\{w_i\} \ . \]
The new plots confirm our previous observations and show how different values of $k$ can be used to modify the shape of the figures in order to highlight their interesting features.

\section{Ensembles with excess covariance}
\label{app:excesscov_APP}

In this section, we derive a Montecarlo procedure to sample random time-series $x_{i,t}$ with excess covariance:
\[ \bra{1}{\hat{M}}\ket{1}=\sum_{i,j}M_{i,j}=N^2c > \left\langle \bra{1}{\hat{M}}\ket{1}\right\rangle_Q \ , \]
where $M_{i,j} = \frac{1}{T}\sum_{t=1}^T x_{i,t}x_{j,t}$ and the matrix elements $x_{i,t}$ are i.i.d.\ random variables drawn from a distribution $Q$. As explained in Sec.\ \ref{sec:marketmode}, our purpose is to understand if the excess covariance is a sufficient hypotesis to explain the emergence of a \emph{market mode} in financial time series, i.e.\ of a principal mode characterized by an anomalous large eigenvalue and an extended eigenvector.
At variance with the prevoius section, here we do not keep fixed the marginal distribution of matrix elements by means of a reshuffling procedure, but rather we allow the elements $x_{i,t}$ to deviate according to their probability distribution $Q$.
The new algorithm is based on the following steps:
\begin{enumerate}
\item Generate a time series $x_{i,t}$ such that the total covariance $\bra{1}{\hat{M}}\ket{1}$ is equal to the desired value $N^2c$.
\item Pick some elements $(x_{i_1,t_1},\dots,x_{i_K,t_K})$ at random (with $K\geq2$).
\item Draw as many random steps $(\epsilon_{i_1,t_1},\dots,\epsilon_{i_K,t_K})$ and add them to the selected elements. The random steps must be choosen such that the final value of the total covariance remains unchanged (this is equivalent to imposing a quadratic constraint on the $\epsilon_{i,t}$'s).
\item Accept the move with probability:
\[ \frac{q(x_{i_1,t_1}+\epsilon_{i_1,t_1})}{q(x_{i_1,t_1})} \times\cdots\times \frac{q(x_{i_K,t_K}+\epsilon_{i_K,t_K})}{q(x_{i_K,t_K})} \ , \]
where $q(x)$ is the p.d.f.\ of the $x_{i,t}$'s, otherwise reject it.
\item Repeat steps (ii)--(iv) until a stationary state is reached.
\end{enumerate}
The above algorithm corresponds to a \emph{micro-canonical} Montecarlo algorithm since, at each step, it keeps $\mathcal{H}=-\bra{1}\hat M\ket{1}$ to a fixed value. It could be tempting to drop the constraints on the random steps $(\epsilon_{i_1,t_1},\dots,\epsilon_{i_K,t_K})$ and use instead a \emph{canonical} algorithm, where the distribution of the matrix $x_{i,t}$ is modified by a factor $e^{-\beta\mathcal{H}}$. The main problem with this choice is that $\mathcal{H}$ is a quadratic function of the $x$'s and so, for $\beta>0$, the new distribution is well-defined only if $q(x)$ dacays at least as fast as a normal distribution (this is not the case for fat-tailed distributions).

\begin{figure}
\includegraphics[width=\textwidth]{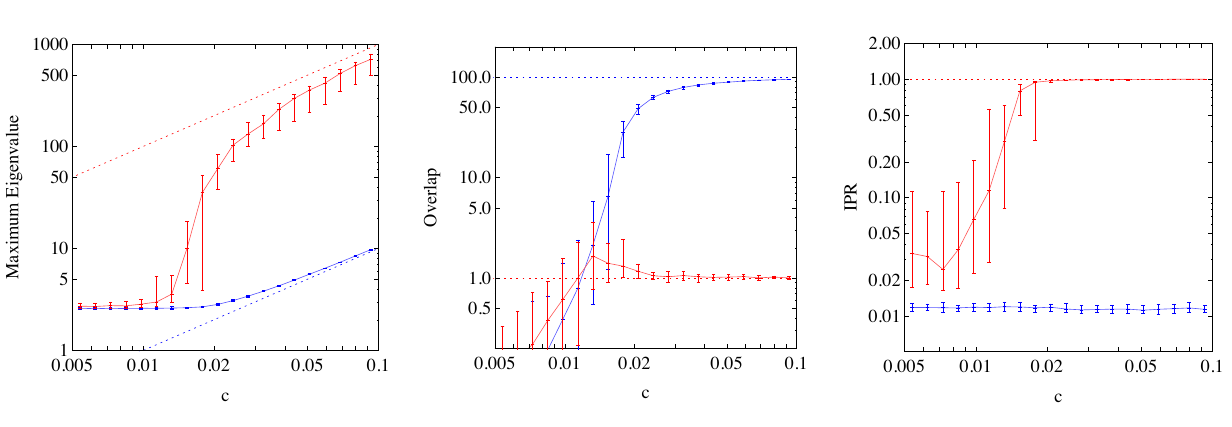}
\caption{Ensembles with excess covariance. From left to right, as a function of $c$, we plot  $\lambda_1$ (left), the overlap $|\braket{1}{\lambda_1}|^2$  (center) and the Inverse Participation Ratio (IPR) of $\braket{x_t}{\lambda_1}$ (right), which measures the degree of concentration of the mode $\ket{\lambda_1}$ over time.
The latter is defined as $\mathrm{IPR}=\sum_t \braket{x_t}{\lambda_1}^4/(\sum_t\braket{x_t}{\lambda_1}^2)^2$, where $\ket{x_t}$ is the vector whose $i^{\rm th}$ component is $x_{i,t}$. We expect $\mathrm{IPR}\sim 1/T$ when $\braket{x_t}{\lambda_1}\simeq\mathrm{const}$ is completely delocalized in $t$ and $\mathrm{IPR}\sim 1$ when $\braket{x_t}{\lambda_1}$ is a completely concentrated on just one value of $t$. In all plots, we report results for the Normal distribution (Blue) and the Student's t-distribution with 4 degrees of freedom (Red).
The results correspond to the financial time series of $N=100$ stocks over $T=250$ days.
}
\label{FIG:MC_A}
\end{figure}

We exploited the above algorithm to simulate the time series of price returns of $N=100$ stocks over $T=250$ days (the size of the simulated time-series has been defined in accordance with the empirical data analyzed in Sec.\ \ref{sec:marketmode}). We made two different choice for the returns' distribution $q(x)$, namely, a normal distribution and a Student-t distribution. The latter is a fat tailed distribution with no rotation invariance and is a good approximation of the actual distribution of returns (see Fig.\ \ref{FIG:PDF}). The results of the algorithm are shown in Fig.\ \ref{FIG:MC_A}. For both the normal and the Students's case, large deviations are concentrated on the largest eigenvalue $\lambda_1$ which separates from the bulk of the other ones, and we find roughly $\bra{1}{\hat{M}}\ket{1} \approx \lambda_1 |\braket{1}{\lambda_1}|^2$. Nevertheless, the way in which the two deviations are realized is deeply different. For the normal distribution we obtain exactly the desired results: the largest eigenvalue scales as $\lambda_1\sim cN$, and the corresponding eigenvector is completely oriendetd towards the market mode, since $|\braket{1}{\lambda_1}|^2\approx N$.
The case of the Student-t distribution, instead, is drastically different. The main eigenvector does not align to $\ket{1}$ and is fully localized. As a result, the overlap $|\braket{1}{\lambda_1}|^2$ converge to 1, and the deviation is realized through the anomalus scaling $\lambda_1\sim cN^2$. Moreover, the localization of $\ket{\lambda_1}$ occurs not only across stocks but also in time (the IPR is very close to 1), and this suggest that large deviations with excess covariance $\bra{1}\hat M\ket{1}$ (as well as with excess volatility ${\rm Tr}~\hat M$) are typically concentrated on one anomalous large return occurring at one specific stock and at one specific day.

In conclusion, the excess covariance is actually enough to justify the emercence of the market mode in financial correlation matrices, but only if the distribution of price returns is thin tailed. This is clearly in contrast with the evidence that the marginal distributions of price returns are power-law. In order to extend such results to fat tailed distributions, we need an additional constraints on the returns' marginal distribution preventing concentration phenomena. Such case has been descussed in Sec.\ \ref{sec:marketmode}.


\section*{References}

\begin{thebibliography}{99}

\bibitem{ellis}
Ellis R S, 1985 {\em Entropy, Large Deviations, and Statistical Mechanics} (New York, NY: Springer)

\bibitem{varadhan}
Varadhan S R S, {\em Large deviations}, 2008 Ann. Prob. {\bf 36} 397

\bibitem{Touchette}
Touchette H, {\em The large deviation approach to statistical mechanics}, 2009 Phys. Rep. {\bf 478} 1

\bibitem{Cov}
Cover T M and Thomas J A, 1991 {\em Elements of Information Theory -- Wiley Series in Telecommunications} (New York, NY: Wiley)

\bibitem{MM}
M\'ezard M and Montanari A, 2009 {\em Information, Physics and Computation} (Oxford: Oxford University Press)

\bibitem{Burda}
Bialas P, Burda Z and Johnston D, {\em Phase diagram of the mean field model of simplicial gravity}, 1999 Nucl. Phys. B {\bf 542} 413

\bibitem{ZRP}
Evans M R and Hanney T, {\em Nonequilibrium statistical mechanics of the zero-range process and related models}, 2005 J. Phys. A {\bf 38} R195

\bibitem{satya}
Majumdar S N, Evans M R and Zia R K P, {\em Nature of the condensate in mass transport models}, 2005 Phys. Rev. Lett. {\bf 94} 180601

\bibitem{satya_2}
Evans M R, Majumdar S N and Zia R K P, {\em Canonical analysis of condensation in factorised steady state}, 2006 J. Stat. Phys. {\bf 123} 357

\bibitem{Embrechts}
Embrechts P, Kl\"uppelberg C and Mikosch T, 1997 {\em Modelling Extremal Events, for Insurance and Finance} (Berlin: Springer)

\bibitem{Wigner}
Wigner E P, {\em On the statistical distribution of the widths and spacings of nuclear resonance levels}, 1951 Proc. Cambridge Philos. Soc. {\bf 47} 790

\bibitem{Wishart}
Wishart J, {\em The generalized product moment distribution in samples from a normal multivariate population}, 1928 Biometrika {\bf 20A} 32

\bibitem{Stanley}  
Gopikrishnan P, Meyer M, Amaral L A N and Stanley H E, {\em Inverse cubic law for the distribution of stock price variations}, 1998 Eur. Phys. J. B {\bf 3} 139

\bibitem{Kirilenko}
Kirilenko A, Kyle A S, Samadi M and Tuzun T, {\em The flash crash: The impact of high frequency trading on an electronic market}, 2011 SSRN: \url{http://ssrn.com/abstract=1686004}

\bibitem{Bormetti}
Bormetti G, Calcagnile L M, Treccani M, Corsi F, Marmi S and F Lillo, {\em Modelling systemic price cojumps with Hawkes factor models}, 2013 \url{http://arxiv.org/abs/1301.6141}

\bibitem{Laloux}
Laloux L, Cizeau P , Bouchaud J-P and Potters M, {\em Noise dressing of financial correlation matrices}, 1999 Phys. Rev. Lett. {\bf 83} 1467

\bibitem{Plerou}
Plerou V, Gopikrishnan P, Rosenow B, Amaral L A N and Stanley H E, {\em Universal and non-universal properties of cross-correlations in financial time series}, 1999 Phys. Rev. Lett. {\bf 83} 1471

\bibitem{Plerou2}
Plerou V, Gopikrishnan P, Rosenow B, Amaral L A N, Guhr T and Stanley H E, {\em Random matrix approach to cross correlations in financial data}, 2002 Phys. Rev. E {\bf 65} 066126

\bibitem{RaffaelliMarsiliPonsot} 
Marsili M, Raffaelli G and Ponsot B, {\em Dynamic instability in generic model of multi-assets markets}, 2009 J. Econ. Dyn. Control {\bf 33} 1170

\bibitem{CAPM}
Fama E F and French K R, {\em The capital asset pricing model: Theory and evidence}, 2003 J. Econ. Perspect. {\bf 18} 25

\bibitem{Gned}
Gnedenko B V, 1998 {\em Theory of Probability} (Boca Raton, FL: CRC Press)

\bibitem{didierbook}
Sornette D, 2004 {\em Critical Phenomena in Natural Sciences} (Heidelberg: Springer)

\bibitem{Zaliapin}
Zaliapin I V, Kagan Y Y and Schoenberg F P, {\em Approximating the distribution of Pareto sums}, 2005 Pure Appl. Geophys. {\bf 162} 1187

\bibitem{BouchaudMezard}
Bouchaud J-P and M\'ezard M, {\em Wealth condensation in a simple model of economy}, 2000 Physica A {\bf 282} 536

\bibitem{evans}
Evans M R and Majumdar S N, {\em Condensation and extreme value statistics}, 2008 J. Stat. Mech. P05004

\bibitem{sf}
Frisch U and Sornette D, {\em Extreme deviations and applications}, 1997 J. Phys. I France {\bf 7} 1155

\bibitem{Szavits-Nossan} 
Szavits-Nossan J, Evans M R and Majumdar S N, {Constraint-driven condensation in large fluctuations of linear statistics}, 2014 Phys. Rev. Lett. {\bf 112} 020602

\bibitem{Johnstone}
Johnstone I, {\em On the distribution of the largest eigenvalue in principal components analysis}, 2001 Ann. Stat. {\bf 29} 295

\bibitem{Tulino}
Tulino A, Verd\'u S, 2004 {\em Random Matrix Theory and Wireless Communications} (Hanover: Now Publishers)

\bibitem{Fukunaga}
Fukunaga K, 1990 {\em Introduction to Statistical Pattern Recognition} (New York, NY: Elsevier)

\bibitem{BouchaudPotters} 
Bouchaud J-P and Potters M, 2000 {\em Theory of Financial Risk and Derivative Pricing} (Cambridge: Cambridge University Press)

\bibitem{EdwardsJones}
Edwards S F and Jones R C, {\em The eigenvalue spectrum of a large symmetric random matrix}, 1976 J. Phys. A: Math. Gen. {\bf 9} 1595

\bibitem{BurdaJurkiewicz}
Burda Z, G\"orlich A, Jarosz A and Jurkiewicz J, {\em Signal and noise in correlation matrix}, 2004 Physica A {\bf 343}, 295

\bibitem{satya3}
Majumdar S N and Schehr G, {\em Top eigenvalue of a random matrix: large deviations and third order phase transition}, 2014 J. Stat. Mech. P01012

\bibitem{Nadal}
Nadal C, Majumdar S N and Vergassola M, {\em Statistical distribution of quantum entanglement for a random bipartite state}, 2011 J. Stat. Phys. {\bf 142} 403

\bibitem{Biroli}
Biroli G, Bouchaud J-P and Potters M, {\em On the top eigenvalue of heavy-tailed random matrices}, 2007 Europhys. Lett. {\bf 78} 10001

\bibitem{Vivo}
Vivo P, Majumdar S N and Bohigas O, {\em Large deviations of the maximum eigenvalue in Wishart random matrices}, 2007 J. Phys. A: Math. Gen. {\bf 40} 4317

\bibitem{satya2}
Majumdar S N and Vergassola M, {\em Large deviations of the maximum eigenvalue for Wishart and Gaussian random matrices}, 2009 Phys. Rev. Lett. {\bf 102} 060601 

\bibitem{Vivo2}
Majumdar S N and Vivo P, {\em Number of relevant directions in Principal Component Analysis and Wishart random matrices}, 2012 Phys. Rev. Lett. {\bf 108} 200601

\bibitem{Nadal_PRE}
Nadal C and Majumdar S N, {\em Nonintersecting Brownian interfaces and Wishart random matrices}, 2009 Phys. Rev. E {\bf 79} 061117

\bibitem{Nadal_PRL}
Nadal C, Majumdar S N and Vergassola M, {\em  Phase transitions in the distribution of bipartite entanglement of a random pure state}, 2010 Phys. Rev. Lett. {\bf 104} 110501

\bibitem{Jorion}
P. Jorion, 2001 {\em Value at Risk: The New Benchmark for Managing Financial Risk} (New York, NY: McGraw-Hill)

\bibitem{Cont}
Cont R, {\em Empirical properties of asset returns: stylized facts and statistical issues}, 2001 Quant. Finance {\bf 1} 223

\bibitem{Fama}
Fama E F, {\em Efficient capital markets: a review of theory and empirical work}, 1970 J. Finance {\bf 25} 383

\bibitem{MRW_bouchaud}
Borland L, Bouchaud J-P, Muzy J-F and Zumbach G. {\em The dynamics of financial markets -- Mandelbrot's multifractal cascades and beyond}, 2005 Wilmott Magazine, page 86

\bibitem{MRW_bacry}
Bacry E, Delour J and Muzy J-F, {\em Multifractal random walk}, 2001 Phys. Rev. E {\bf 64} 026103

\bibitem{Taylor}
Taylor S, 1986 {\em Modelling Financial Time Series} (New York, NY: Wiley)

\bibitem{Krugman} 
Krugman P, 1995 {\em Development, Geography, and Economic Theory} (Cambridge, MA: The MIT Press)

\bibitem{Gould} 
Gould S J and Eldredge J, {\em Punctuated equilibria: the tempo and mode of evolution reconsidered}, 1977 Paleobiology {\bf 3} 115

\bibitem{DragonKings} 
Sornette D, {\em Dragon-Kings, Black Swans and the Prediction of Crises}, 2009 International Journal of Terraspace Science and Engineering {\bf 2} 1

\end{thebibliography}
\end{document}